\documentclass[showpacs,prl,floatfix,preprintnumbers,twocolumn]{revtex4}
\usepackage{graphicx,url,amssymb,amsmath,longtable,rotating,color,psfig,units,wasysym,subfigure,epsfig,listings,bm}

\begin{document}
\title{A Model of the Stochastic Gravitational-Wave Background due to Core Collapse to Black Holes}
\author{K. Crocker$^{a}$, V. Mandic$^a$\footnote{mandic@physics.umn.edu}, T. Regimbau$^b$, K. Belczynski$^{c}$, W. Gladysz$^c$, K. Olive$^{d,a}$, T. Prestegard$^a$, E. Vangioni$^e$}
\affiliation{$^a$School of Physics and Astronomy, University of Minnesota, Minneapolis, MN 55455, USA\\
$^b$Departement Artemis, Observatoire de la C\^ote d'Azur, CNRS, F-06304 Nice,  France\\
$^c$ Astronomical Observatory, University of Warsaw, Al. Ujazdowskie 4, 00-478 Warsaw, Poland\\
$^d$ William I. Fine Theoretical Physics Institute, University of Minnesota, Minneapolis, MN 55455, USA\\
$^e$ Sorbonne Universit\'es, UPMC Univ. Paris 6 et CNRS, UMR 7095, Institut d’Astrophysique de Paris, 98 bis bd Arago, 75014 Paris, France}
\date{\today}

\begin{abstract}
  Superposition of gravitational waves generated by astrophysical sources is expected to give rise to the stochastic gravitational-wave background. We focus on the background generated by the ring-down of black holes produced in the stellar core collapse events across the universe. We systematically study the parameter space in this model, including the most recent information about the star formation rate and about the population of black holes as a function of redshift and of metallicity. We investigate the accessibility of this gravitational wave background to the upcoming gravitational-wave detectors, such as Advanced LIGO and Einstein Telescope.
\end{abstract}

\pacs{95.85.Sz, 97.60.Jd, 04.25.dg, 98.80.Cq}
\maketitle

\section{Introduction}
The stochastic gravitational-wave background (SGWB) is expected to arise from the superposition of gravitational waves (GWs) from many uncorrelated and unresolved sources.
Numerous cosmological and astrophysical models have been proposed.
Cosmological models include inflationary models~\cite{grishchuk,starob,eastherlim,peloso}, models based on cosmic (super)strings \cite{caldwellallen,DV1,DV2,cosmstrpaper,olmez1,olmez2}, and models of alternative cosmologies \cite{PBB1,PBBpaper}.
Astrophysical models (see \cite{regimbau} for a review) integrate contributions from astrophysical objects across the universe including compact binary coalescences (CBC) of  binary neutron stars (BNS) or binary black holes (BBH) \cite{phinney,kosenko,2006ApJ...642..455R,zhu,2011PhRvD..84h4004R,2011PhRvD..84l4037M,StochCBC,2013MNRAS.431..882Z,2013MPLA...2850174E,2015A&A...574A..58K,2015MNRAS.449.2700E}, rotating neutron stars (NSs) \cite{RegPac,owen,1999MNRAS.303..258F,barmodes1,barmodes2,barmodes3,2004MNRAS.351.1237H,2011ApJ...729...59Z,2012PhRvD..86j4007R,2013PhRvD..87f3004L}, magnetars \cite{cutler,2006A&A...447....1R,RegMan,2011MNRAS.411.2549M,2011MNRAS.410.2123H,2013PhRvD..87d2002W,2013PhRvD..87f3004L}, the first stars \cite{firststars}, and white dwarf binaries~\cite{phinney_whitedwarfs}.

Several searches for the isotropic \cite{S3stoch,S4stoch,S5stoch,S6stoch,H1H2stoch} and anisotropic SGWB \cite{S4radiometer,S5SPH} have been conducted using data acquired by the first generation interferometric gravitational-wave detectors LIGO \cite{LIGOS1,LIGOS5} and Virgo \cite{Virgo1}. These searches have established upper limits on the energy density in the SGWB, and have started to constrain some of the proposed models \cite{cosmstrpaper,StochCBC,paramest,parviol}. The second generation of gravitational-wave detectors is currently being commissioned, including Advanced LIGO (aLIGO) \cite{aLIGO2}, Advanced Virgo \cite{aVirgo}, GEO-HF \cite{GEOHF,GEOHF2}, and KAGRA \cite{KAGRA1,KAGRA2}. These detectors are expected to produce first data in 2015, and their strain sensitivity is expected to be 10$\times$ better relative to the first generation detectors. The third generation gravitational-wave detectors are also being conceptualized, such as the Einstein Telescope for which the design study was completed in Europe \cite{ET}.

One of the promising sources of gravitational waves, potentially detectable by the second and third generation detectors, is the stellar core collapse process.
The physics of core collapse is complex and it is expected to produce gravitational waves via several mechanisms \cite{OttPRL,ott2013,muller2013}. While the full three-dimensional simulations that include all relevant processes are yet to be made, predictions have already been made about the gravitational-wave signals emitted during core collapse \cite{OttPRL,ott2013,muller2013}. These predictions have then been used to make estimates of the corresponding SGWB due to both standard and early (population III) stars \cite{ferrari_ccbh,2000PhRvD..61l4015D,2002MNRAS.330..651D,araujo,2004MNRAS.348.1373D,2004CQGra..21S.545D,buonanno,coward,2010MNRAS.409L.132Z,marassi_cc,firststars,pacucci}. These estimates are necessarily only approximate, both because it is currently not well understood how the gravitational-wave signal depends on the progenitor stellar parameters such as the mass or spin, and because the rate of core collapse events is uncertain.

In this paper, we focus on the ringdown of the black hole (BH) following the core collapse. The GW spectrum emitted by this process is relatively well understood, since most of the energy is dissipated via the ringdown of the $l=2$ dominant quasi-normal mode \cite{echeveria,regimbau}, as is also confirmed in simulations \cite{SP1,araujo}. The SGWB due to this source was first examined by Ferrari et al \cite{ferrari_ccbh}. We revisit this model of SGWB, taking into account the most recent information about the star formation rate, about the population of black holes, and about the effects due to varying metallicity. We also investigate the accessibility of the model to the upcoming second and third generation GW detectors. In Section 2 we present the general aspects of the calculation of the astrophysical SGWB, and we discuss the star formation rate used in this study. In Section 3 we discuss the SGWB due to the core collapse to black holes (Model 1)---this has been discussed in the literature but we revisit it here using the latest star formation rate. In Section 4, we discuss Model 2, which builds on the Model 1 by including the effects of metallicity. In Section 5, we discuss Model 3 which is based on a Monte Carlo simulation of the SGWB due to core collapse to black holes, taking into account the StarTrack \cite{startrack} numerical simulation of the black hole population. We summarize our results in Section 6.

\section{Astrophysical Gravitational-wave Background}
To compute the SGWB due to astrophysical GW sources, we follow the formalism used in \cite{regimbau,RegMan,StochCBC}. In particular, we define the normalized energy density in gravitational waves:
\begin{eqnarray}
\Omega_{\rm GW}(f) & = & \frac{1}{\rho_c } \frac{d\rho_{\rm GW}}{d \ln f},
\end{eqnarray}
where $\rho_{\rm GW}(f)$ is the energy density in gravitational waves at the observed frequency $f$, and $\rho_c = \frac{3 H_0^2 c^2}{8\pi G}$ is the critical energy density needed to close the universe. Here, $H_0$ is the present value of the Hubble parameter, taken to be 68 km/s/Mpc, $G$ is the Newton's constant, and $c$ is the speed of light. The energy density can then be rewritten in terms of the integrated flux $F(f)$ over redshift $z$:
\begin{eqnarray}
\Omega_{\rm GW}(f) & = & \frac{f}{\rho_c c} F(f) \nonumber \\
& = & \frac{f}{\rho_c c} \int dz \frac{R_z(z)}{4\pi r^2(z)} \; \frac{dE_{\rm GW}}{df_e}.
\end{eqnarray}
Here, $r(z)$ is the proper distance, and $dE_{\rm GW}/df$ is the energy spectrum emitted by a single astrophysical source as a function of the emitted frequency $f_e$ in the source's frame, $f_e = f(1+z)$. We will consider three different models for $dE_{\rm GW}/df_e$ below. The rate of astrophysical sources as a function of redshift, $R_z(z)$, can be written in terms of the rate of sources per comoving volume $R_V(z)$:
\begin{eqnarray}
R_z(z) & = & R_V(z) \frac{dV}{dz} \nonumber \\
& = & R_V(z) \frac{4\pi c}{H_0} \frac{r^2(z)}{E(\Omega_M,\Omega_{\Lambda},z),}
\end{eqnarray}
where $E(\Omega_M,\Omega_{\Lambda},z)$ captures the dependence of the comoving volume on redshift:
\begin{eqnarray}
E(\Omega_M,\Omega_{\Lambda},z) = \sqrt{\Omega_M(1+z)^3 + \Omega_{\Lambda}},
\end{eqnarray}
with $\Omega_M = 0.3$ and $\Omega_{\Lambda} = 0.7$ corresponding to the energy density in matter and dark energy respectively. Finally, the rate of astrophysical sources can be related to the star formation rate (SFR) $R_*(z)$:
\begin{equation}
R_V(z) = \lambda_{\rm BH} \frac{R_*(z)}{1+z},
\label{eqrv}
\end{equation}
where $\lambda_{BH}$ captures the mass fraction of matter that ends up in progenitors of black holes. Bringing the above together, we get
\begin{eqnarray}
\Omega_{\rm GW}(f) & = & \frac{8\pi Gf\lambda_{\rm BH}}{3H_0^3 c^2} \int dz \frac{ R_*(z)}{(1+z)E(\Omega_M,\Omega_{\Lambda},z)} \; \frac{dE_{\rm GW}}{df_e}.\nonumber \\
&&
\end{eqnarray}
The star formation rate and its dependence on redshift have been studied extensively in the literature, and multiple functional forms have been proposed \cite{lilly,hopkins,fardal,wilkins,nagamine,springel}. Historically, most of the star formation rate estimates have been based on luminosity measurements.
We use the SFR derived in Behroozi et al. \cite{behroozi} and Oesch et al. \cite{oesch1,oesch2}. While luminosity measurements at redshifts up to $\sim 2$ are relatively well understood, measurements at high redshifts (up to $z \sim 11$) are subject to uncertainties due to the extinction by dust and due to the fact that early star formation takes place in faint galaxies that may be missed in magnitude-limited surveys. An alternative to luminosity measurements at high redshift is to use the rate of Gamma Ray Bursts (GRBs) \cite{robertsonellis,wang,kistler}, which typically results in a slower fall-off of the star formation rate at high redshifts. We use the GRB rate of \cite{kistler} based on the normalization described in \cite{trenti,behroozisilk}.

\begin{figure}
  \psfig{file=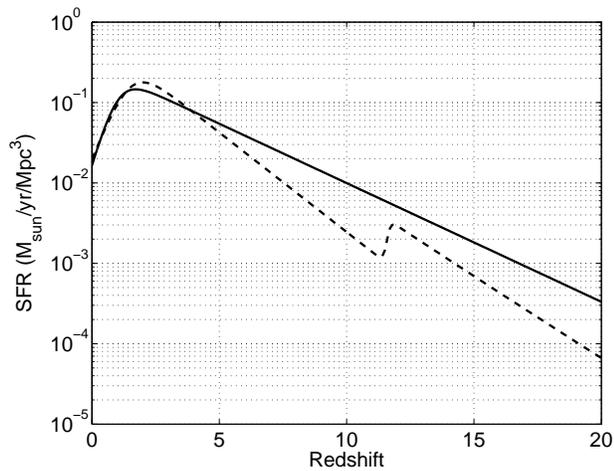, width=3.2in}
  \caption{Comparison of the GRB-based (solid) and luminosity-based (dashed) models of star formation rate.
      }
  \label{sfr}
\end{figure}

As argued in \cite{vangioni}, the choice of the star formation rate has direct implications for the chemical and reionization history of the universe. The analysis presented in \cite{vangioni} considered both the GRB and the luminosity data and arrived at two models of star formation rate that are consistent with the available metallicity data as well as with the reionization redshift and the optical depth measurement by WMAP \cite{WMAP}. Both models use the Springel \& Hernquist functional form \cite{springel}:
\begin{eqnarray}
R_*(z) = \nu \; \frac{p e^{q(z-z_m)}}{p-q+q e^{p(z-z_m)}}
\end{eqnarray}
with the following parameters \cite{vangioni}:
\begin{itemize}
\item GRB-based model \cite{kistler,trenti,behroozisilk}: $\nu= 0.146 \;  M_{\odot}/{\rm yr}/{\rm Mpc}^3 $, $z_m = 1.72$, $p = 2.80$, and $q = 2.46$.
\item Luminosity-based model \cite{behroozi,oesch1,oesch2}: containing the normal mode stars described by $\nu = 0.178 \;  M_{\odot}/{\rm yr}/{\rm Mpc}^3 $, $z_m = 2.00$, $p = 2.37$, and $q = 1.80$, {\it and} population III stars described by $\nu = 0.00218 \;  M_{\odot}/{\rm yr}/{\rm Mpc}^3 $, $z_m = 11.87$, $p = 13.81$, and $q = 13.36$.
\end{itemize}
Furthermore, both models use the same Salpeter initial mass function (IMF) $\phi(m) = N m^{-2.35}$, where $N$ is a normalization constant defined such that
\begin{equation}
\int_{m_1}^{m_2} \phi(m) m dm = 1.
\end{equation}
Note that the normalization constant depends very weakly on the upper limit of this integral, so we set $m_2 = \infty$.
For the normal-mode stars we choose the lower limit of the integral to be $m_1 = 0.1 M_{\odot}$. Since population III stars are expected to be heavier, we choose $m_1 = 36 M_{\odot}$ for this population \cite{vangioni}.
Figure \ref{sfr} compares these two models of star formation rate as a function of redshift.
We will compare our results for these two models of star formation rate. Note that these two models predict substantially less star formation at high redshifts as compared with some of the models used in the past to estimate the stochastic gravitational-wave backgrounds (e.g. \cite{firststars}).

\section{Model 1}
We first consider the model of SGWB due to core collapse to black holes that was first proposed by Ferrari et al \cite{ferrari_ccbh}. Numerical simulations \cite{SP1,araujo} have shown that most of the energy is dissipated via the ringdown of the $l=2$ dominant quasi-normal mode, whose frequency is given by \cite{echeveria,regimbau}:
\begin{eqnarray}
\nu_* (m,a,\alpha) & = & \frac{\Delta(a)}{\alpha m} \\
\Delta(a) & = & \frac{c^3}{2\pi G} (1 - 0.63 (1-a)^{0.3})
\label{eqnu}
\end{eqnarray}
where the mass of the black hole $M$ is assumed to be a fraction $\alpha$ of the mass of the progenitor $m$ ($M=\alpha m$), and
$a$ is the dimensionless spin factor ranging from 0 for a Schwarzschild black hole to 1 in the extreme Kerr limit. The energy spectrum of the single source is therefore
\begin{equation}
\frac{dE_{\rm GW}}{df_e} = \epsilon \alpha mc^2 \delta(f_e - \nu_{*}(m,a,\alpha)),
\label{eqenergy}
\end{equation}
where $\epsilon$ is the efficiency of GW production, acting as a scaling parameter of the GW energy spectrum. Hence,
\begin{eqnarray}
\Omega_{\rm GW}(f) & = & \frac{8\pi Gf\epsilon\alpha}{3H_0^3} \int dz \int_{m_{\rm min}}^{m_{\rm max}} dm  \\
&& \frac{ R_*(z) \; \phi(m) \; m \; \delta(f(1+z) - \nu_{*}(m,a,\alpha))}{(1+z) E(\Omega_M,\Omega_{\Lambda},z)}. \nonumber
\end{eqnarray}
We can use the Dirac delta function to evaluate the integral over redshift, resulting in:
\begin{eqnarray}
\Omega_{\rm GW}(f) & = & \frac{8\pi G\epsilon\alpha}{3H_0^3} \int_{m_{\rm min}}^{m_{\rm max}} dm \frac{ R_*(z')\;  \phi(m) \; m }{(1+z') \; E(\Omega_M,\Omega_{\Lambda},z')}, \nonumber \\
&&
\label{omega_zint}
\end{eqnarray}
where
\begin{equation}
z' = \frac{\nu_*(m,a,\alpha)}{f} - 1.
\end{equation}
Note that since $z'>0$ necessarily, the integral over $m$ includes only the values of $m$ for which $\nu_* (m,a,\alpha) > f$.

The range of integration in Eq. \ref{omega_zint} is defined by some minimum and maximum stellar masses that are expected to act as black hole progenitors. The lower end of this range, $m_{\rm min}$ affects only the high-frequency end of the GW spectrum, above $\sim 200$ Hz. Since most of the sensitivity of GW detectors to this SGWB comes from the low-frequency end of the spectrum, $m_{\rm min}$ does not strongly affect the accessibility of the model to the GW detectors, and we will fix it to $40 M_{\odot}$. Also note that $\alpha m$ must be larger than $\sim 4 M_{\odot}$, otherwise the black hole will not be produced. We verified that the GW spectrum is not very sensitive to the cutoff on $\alpha m$, for example setting the cutoff at $3 M_{\odot}$ has negligible impact on $\Omega_{\rm GW}$.

The high end of the integration range $m_{\rm max}$ affects the low-frequency end of the spectrum and could therefore have a significant impact on the detectability of the model by GW detectors. Furthermore, there is currently much uncertainty in the largest mass of the black hole progenitor stars---for example, stars as massive as $\sim 200-300 M_{\odot}$ stars we recently reported in the R136 star cluster in the LMC \cite{Crowther2010}. To capture this uncertainty, we will repeat our analysis for two values of $m_{\rm max} = 100 M_{\odot}$ and $500 M_{\odot}$ to illustrate the importance of this high-mass cutoff.

\begin{figure}
\includegraphics[width=3.2in]{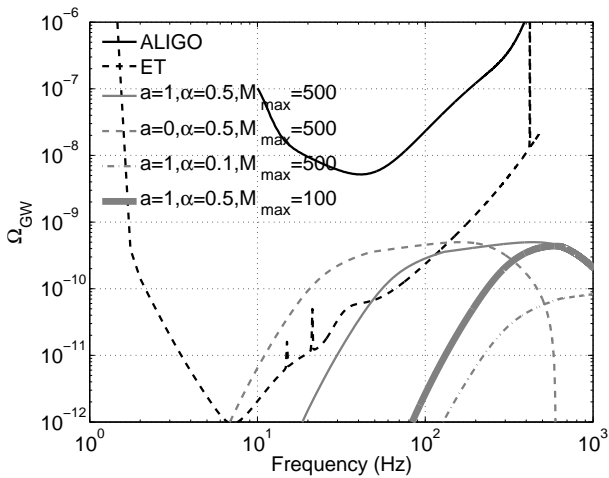}
  \caption{The expected sensitivities of Advanced LIGO \cite{aLIGOsens} and Einstein Telescope \cite{ET} detectors (assuming 1 year exposure) are shown in comparison with several examples of GW spectra obtained using Model 1 with $\epsilon = 10^{-5}$ and with the GRB-based model of star formation rate. While we assume co-located Advanced LIGO detectors in this study, their sensitivity to SGWB will be very similar to the sensitivity of the complete second-generation GW detector network, including Advanced LIGO, Advanced Virgo, GEO-HF, and KAGRA.     }
  \label{sensitivity}
\end{figure}

The free parameters of the model are therefore $\epsilon, \alpha,$ and $a$. Figure \ref{sensitivity} shows example spectra for several choices of parameter values, in comparison with the expected sensitivity of Advanced LIGO and Einstein Telescope detectors. The $\epsilon$ parameter is simply a scaling factor, and $a$ effectively shifts the GW spectrum in frequency: low values of $a$ shift the spectrum to lower frequencies, at which the GW detectors are more sensitive. The parameter $\alpha$ has a more complex impact on the GW spectrum: low values of $\alpha$ shift the spectrum to higher frequencies and reduce its amplitude, both of which reduce the accessibility of the model to GW detectors. Finally, Figure \ref{sensitivity} shows the impact of the upper cutoff on the black hole progenitor mass, $m_{\rm max}$: increasing this cutoff from $100 M_{\odot}$ to $500 M_{\odot}$ extends the GW spectrum to $\sim 5\times$ lower frequencies, as expected based on Equations \ref{eqnu} and \ref{eqenergy}.

We scan this parameter space, restricting $0<a<1$ and $0.1<\alpha<0.5$ (motivated by simulations \cite{belczynski}, as discussed further below). For each point in this parameter space, we compute the spectrum $\Omega_{\rm GW}(f)$ and compare it to the sensitivities of the Advanced LIGO and ET detectors, computing the likelihood function:
\begin{eqnarray}
L \propto \prod_i e^{-(Y_i - \Omega_{{\rm GW},i}(\epsilon,a,\alpha))^2 / 2 \sigma_i^2}
\label{likelihood_eq}
\end{eqnarray}
where the index $i$ runs over frequency bins, $Y_i$ is the expected measurement of the GW energy density in the bin $i$, $\sigma_i$ is the corresponding measurement error, and $\Omega_{{\rm GW},i}(\epsilon,a,\alpha)$ is the value of the energy density in the bin $i$ for the given free parameters $\epsilon,a$, and $\alpha$. For projecting the future experimental sensitivities we set $Y_i=0$. In order to determine the accessibility of this 3-dimensional parameter space to future detectors, we marginalize (integrate) the likelihood function over one of the parameters, and then compute the 95\% confidence level contours in the plane of the remaining two parameters.

\begin{figure*}[!t]
   \begin{tabular}{cc}
\includegraphics[width=3.2in]{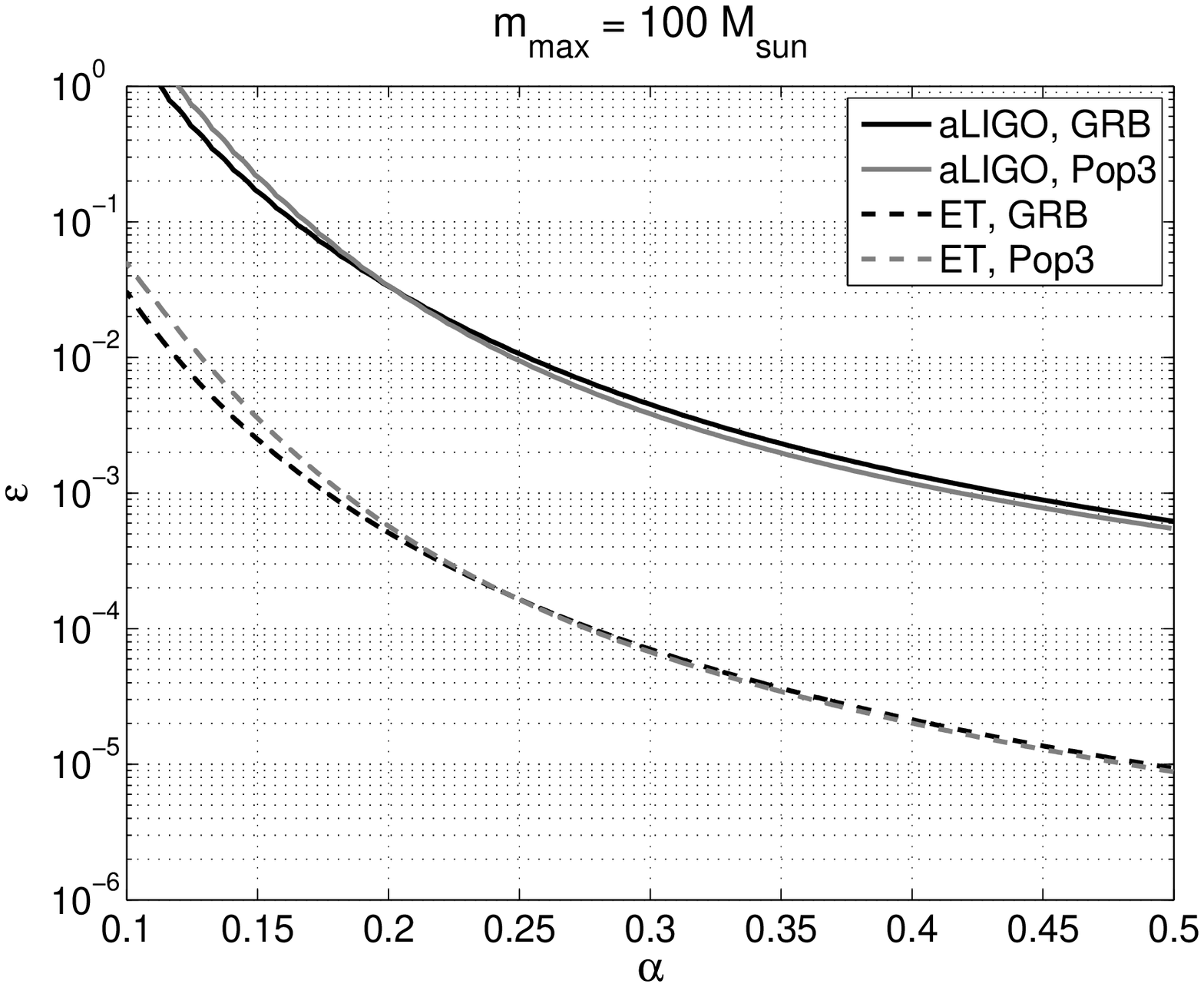} &
\includegraphics[width=3.2in]{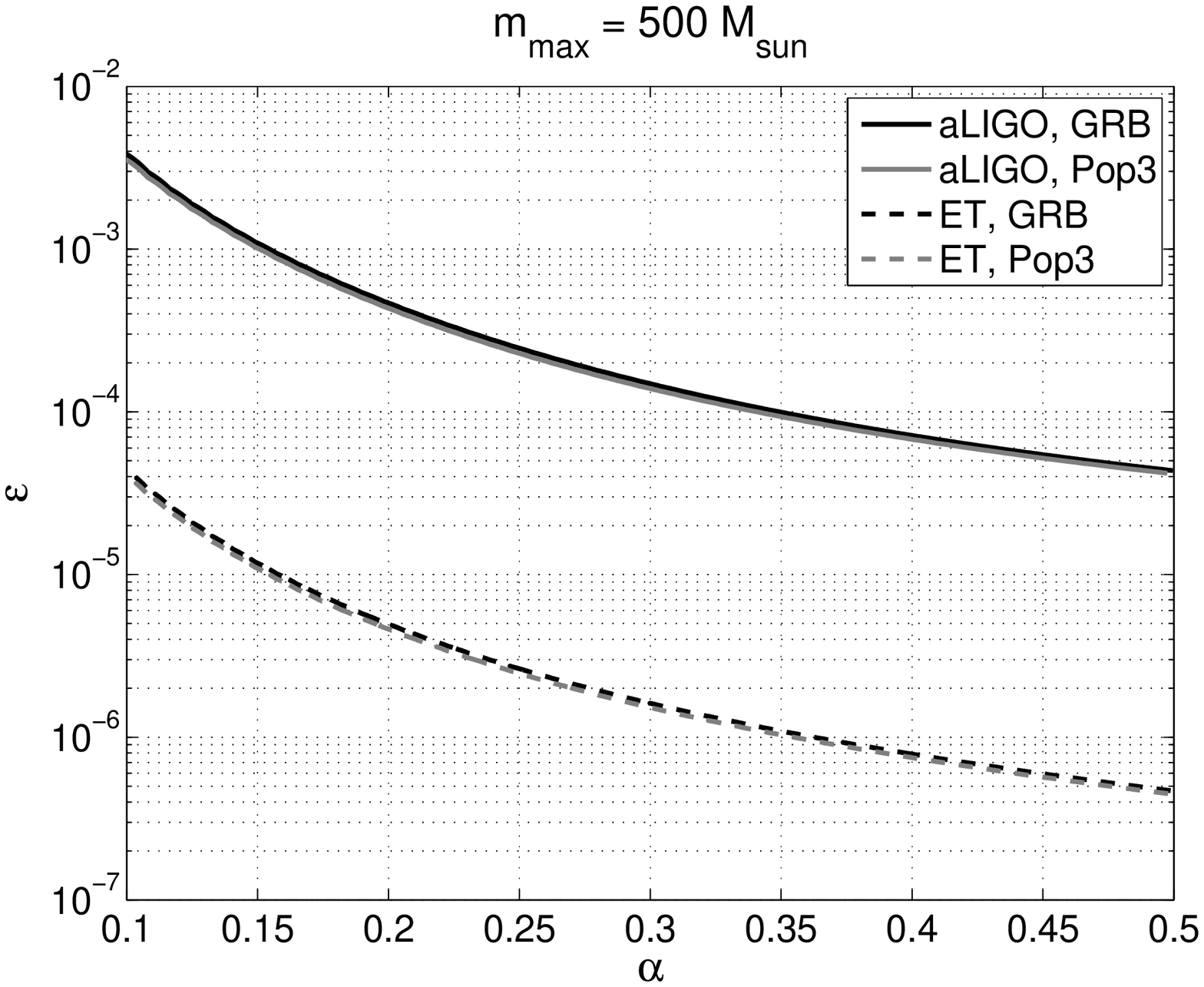} \\
\includegraphics[width=3.2in]{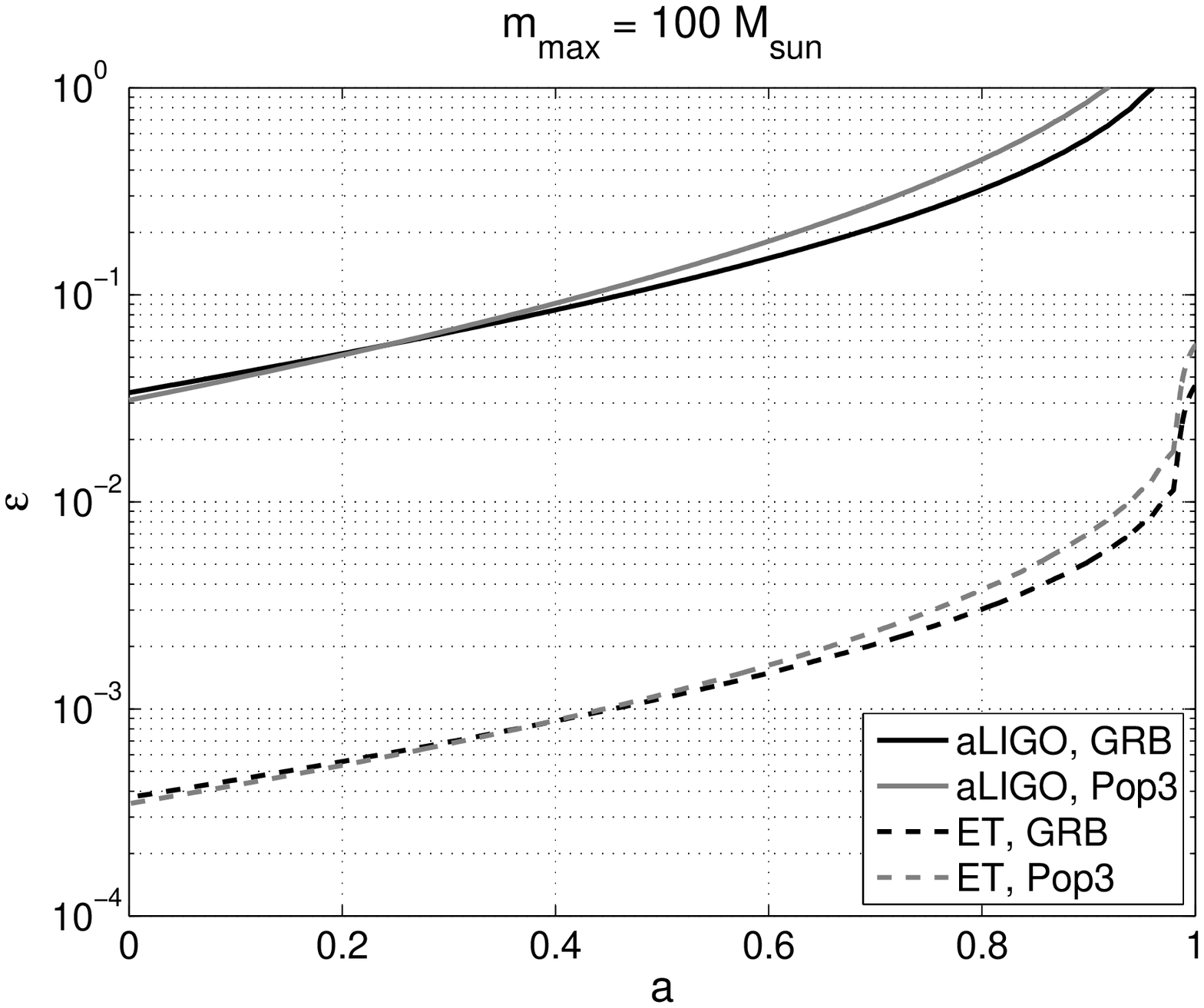} &
\includegraphics[width=3.2in]{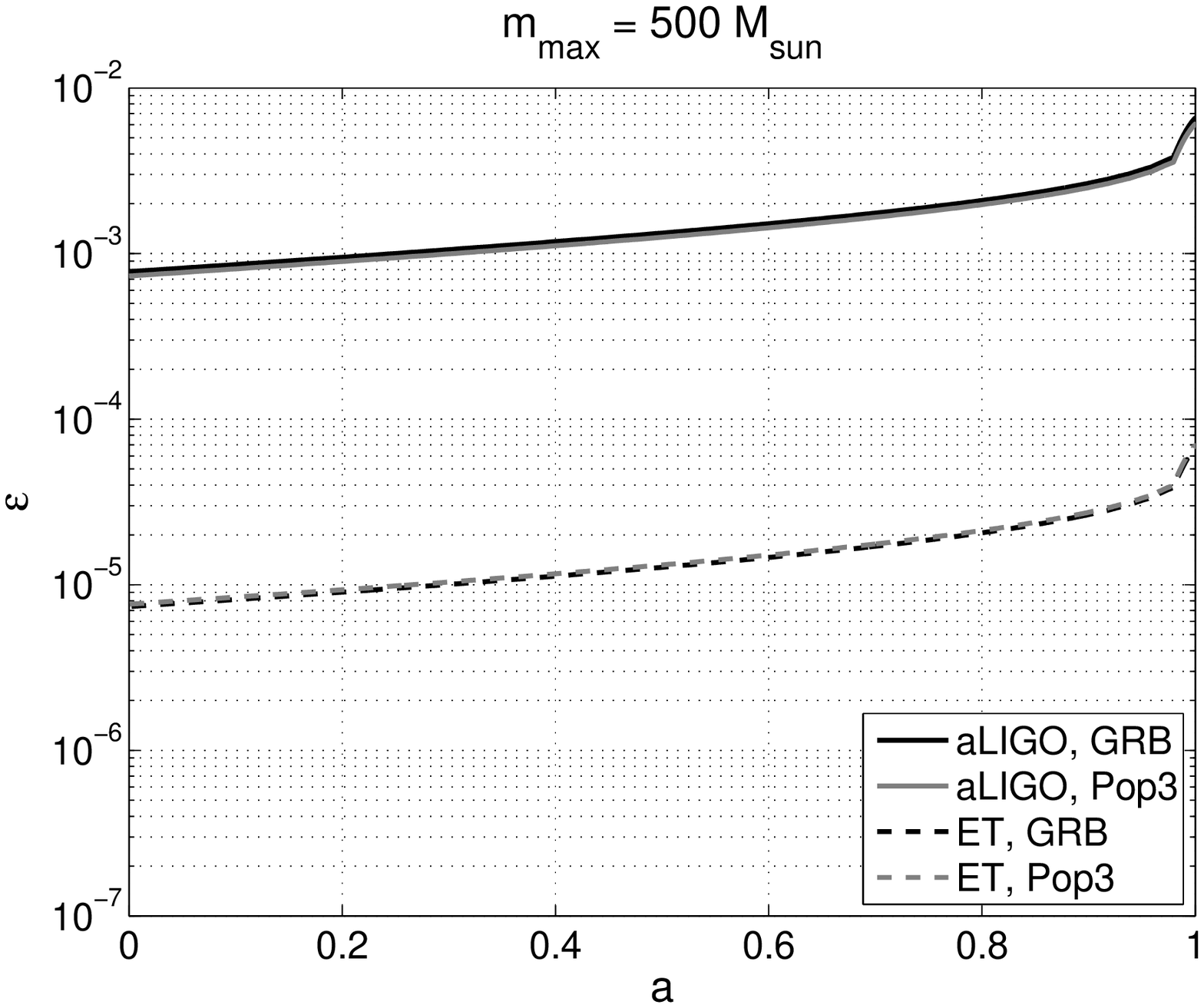} \\
  \end{tabular}
  \caption{Top left: 95\% confidence sensitivity contours of the Advanced LIGO and Einstein Telescope and for two models of star formation rate are shown in the $\epsilon-\alpha$ plane for Model 1, after marginalizing over the parameter $a$, and assuming $m_{\rm max}=100 M_{\odot}$. Top right: Same as top left but for $m_{\rm max}=500 M_{\odot}$. Bottom left: 95\% confidence sensitivity contours of the Advanced LIGO and Einstein Telescope and for two models of star formation rate are shown in the $\epsilon-a$ plane for Model 1, after marginalizing over the parameter $\alpha$, and assuming $m_{\rm max}=100 M_{\odot}$. Bottom right: Same as bottom left but for $m_{\rm max}=500 M_{\odot}$. Note the significant change of vertical scale from models with $m_{\rm max} = 100 M_{\odot}$ to models with $m_{\rm max} = 500 M_{\odot}$.
      }
  \label{model1_contours}
\end{figure*}

Figure \ref{model1_contours} shows the contours computed in this way for the $\epsilon-\alpha$ and $\epsilon-a$ planes (top and bottom rows respectively) and for two choices of $m_{\rm max}=100 M_{\odot}$ and $500 M_{\odot}$ (left and right columns respectively). Curves for both Advanced LIGO and Einstein Telescope, and for both the GRB-based and the luminosity-based star formation rates are shown. The curves in the $\epsilon-\alpha$ plane are decreasing with $\alpha$, which is a consequence of the fact that increasing $\alpha$ pushes the GW spectrum to lower frequencies (c.f. Figure \ref{sensitivity}), hence making the model more detectable by the GW detectors. Similarly, increasing the value of $a$ pushes the GW spectrum to higher frequencies (c.f. Figure \ref{sensitivity}), making the model less accessible to GW detectors and causing the increasing trend (with $a$) of the curves in the $\epsilon-a$ plane. The upper cutoff on the black hole progenitors mass also has a significant impact on the accessibility of these models. As seen by comparing the two columns of Figure \ref{model1_contours}, changing $m_{\rm max}$ from $100 M_{\odot}$ to $500 M_{\odot}$ lowers the sensitivity curves in the $\epsilon$ parameter by 2-3 orders of magnitude. We verified that increasing $m_{\rm max}$ to $1000 M_{\odot}$ further lowers the sensitivity curves by another factor of 2.

The two star formation rates yield nearly identical predictions, which is the consequence of the fact that the dominant contribution to $\Omega_{\rm GW}$ comes from redshifts region of 1-2, in which the two models of star formation rate agree well. It is also evident that the Einstein Telescope will provide a substantially better probe of this model than the second generation detectors.

Finally, we note that the expected value of $\epsilon$ in this mechanism for GW production is uncertain. While some of the past literature considers $\epsilon$ as high as 0.01 \cite{kobayashi}, previous simulations by Stark and Piran \cite{SP1} gave an upper limit of $\epsilon \sim 7 \times 10^{-4}$ for an axisymmetric collapse. Accounting for more realistic scenarios, in particular the pressure reduction that triggers the collapse, leads to $\epsilon \sim 10^{-7}-10^{-6}$ \cite{baiotti}. These values are clearly out of reach of the second-generation detectors as shown in Figure \ref{model1_contours}, but may be within reach of Einstein Telescope, especially in the case if massive black hole progenitors exist (corresponding to the case $m_{\rm max} = 500 M_{\odot}$).

\section{Model 2}
\begin{figure}
\includegraphics[width=3.2in]{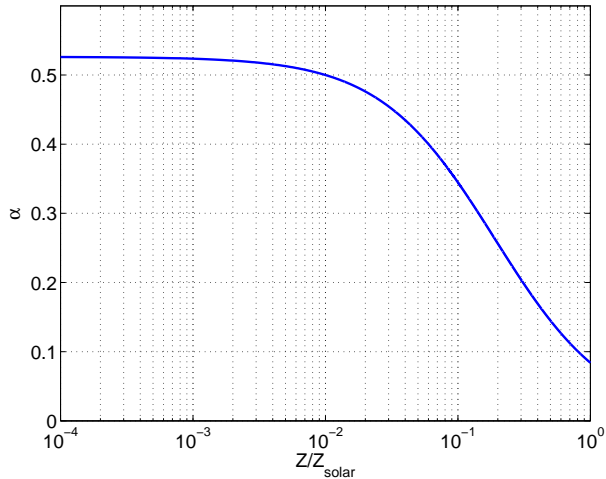}
  \caption{Dependence of $\alpha$ (the fraction of the progenitor mass that remains in the black hole) on metallicity, extracted from the simulation results of \cite{belczynski}. }
  \label{alphaZ}
\end{figure}

We now modify the model discussed above to take into account the effects of metallicity $Z$. Metallicity has been shown to impact the maximum masses of the black hole progenitors as well as the fraction of the progenitor mass that remains in the black hole $\alpha(Z)$ \cite{belczynski}. In particular, simulation results described in \cite{belczynski} provide an explicit formula for the maximum black hole mass as a function of metallicity (Eq. 11 of \cite{belczynski}). Furthermore, this study shows the dependence of the remnant mass $M_{BH}$ on the progenitor mass $m$ for several values of metallicity - we model this dependence as linear, $M_{BH} = \alpha(Z) m$, extracting the value of $\alpha$ for several metallicity values, and then interpolating as necessary. For the sake of simplicity, we ignore the dependence of $\alpha$ on the progenitor mass $m$. The resulting curve for the average $\alpha(Z)$ is shown in Figure \ref{alphaZ}. Finally, we note that the minimum mass of black hole progenitors also depends on the metallicity, as we will see in Model 3---for Model 2, however, we will assume that $m_{\rm min} (Z) = 20 M_{\odot}$, independent of the metallicity. Also, as in Model 1, we require $\alpha m > 4 M_{\odot}$ in order for the black hole to be produced.
\begin{figure}
\includegraphics[width=3.2in]{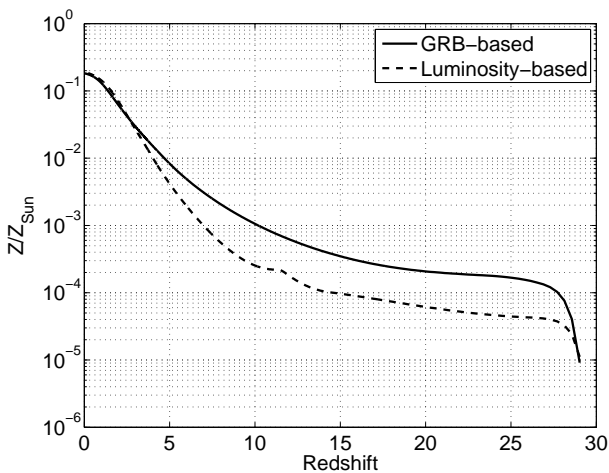}
  \caption{Evolution of metallicity is shown for the two star formation rate models considered in this paper. Note that $Z_{\rm Sun} = 0.02$.      }
  \label{metallicity}
\end{figure}

Metallicity is a function of redshift $Z(z)$, and can be estimated using the chemical evolution code for the chosen star formation rate \cite{vangioni}. Figure \ref{metallicity} shows the mean metallicity evolution for the two star formation rates considered in this paper. For any given redshift, we assume that the metallicity values range within a factor of 3 from the mean metallicity, with a uniform (flat) distribution denoted by $\psi(Z;z)$, and we average over it. We can therefore rewrite the equation for $\Omega_{\rm GW}(f)$:
\begin{eqnarray}
&&\Omega_{\rm GW}(f) = \frac{8\pi Gf\epsilon}{3H_0^3} \int dz \frac{ R_*(z)}{(1+z) E(\Omega_M,\Omega_{\Lambda},z)}  \nonumber \\
&& \int dZ \psi(Z;z) \alpha(Z)  \\
&& \int_{m_{\rm min}(Z)}^{m_{\rm max}(Z)} dm \;  \phi(m) \; m \; \delta(f(1+z) - \nu_{*}(m,a,\alpha)). \nonumber
\end{eqnarray}
We first solve the integral over the progenitor mass using the Dirac delta function:
\begin{eqnarray}
\delta(f(1+z) - \nu_{*}(m,a,\alpha)) & = & \frac{\delta(m-m_0) \alpha(Z) mm_0}{\Delta(a)} \\
m_0 & = & \frac{\Delta(a)}{f(1+z)\alpha(Z)}.
\end{eqnarray}
Clearly, the following condition must be satisfied for a non-zero signal: $m_{\rm min} (Z) < m_0 < m_{\rm max} (Z) $. This translates into
\begin{eqnarray}
\frac{\Delta(a)}{m_{\rm min}(Z) (1+z) \alpha(Z)} > & f & > \frac{\Delta(a)}{m_{\rm max}(Z) (1+z) \alpha(Z)} \nonumber \\
f_{\rm max}(Z,a,z) > & f & > f_{\rm min}(Z,a,z)
\label{freqcond}
\end{eqnarray}
where in the last line we have defined $f_{\rm max}$ and $f_{\rm min}$. Then,
\begin{eqnarray}
&& \Omega_{\rm GW}(f) = \frac{8\pi Gf\epsilon}{3H_0^3} \int dz \frac{ R_*(z)}{(1+z) E(\Omega_M,\Omega_{\Lambda},z)}  \nonumber \\
&& \int dZ \psi(Z;z) \alpha(Z)  N m_0^{0.65} \frac{\alpha(Z)}{\Delta(a)} \nonumber \\
&& \Theta(f-f_{\rm min}) \Theta(f_{\rm max}-f) \\
& = & \frac{8\pi Gf^{0.35} \epsilon}{3H_0^3 \Delta(a)^{0.35}} \int dz \frac{ R_*(z)}{(1+z)^{1.65} E(\Omega_M,\Omega_{\Lambda},z) }  \nonumber \\
&& \int dZ \psi(Z;z) \alpha^{1.35}(Z) N \Theta(f-f_{\rm min}) \Theta(f_{\rm max}-f)  \nonumber
\end{eqnarray}
If the condition in Eq. \ref{freqcond} is not satisfied, the $m$-integral is simply zero. Note that $N$ is the normalization of the initial mass function.

Finally, if we ignore the metallicity dependence, the equation simplifies to:
\begin{eqnarray}
\Omega_{\rm GW}(f) & = & \frac{8\pi Gf^{0.35} \epsilon \alpha^{1.35} N}{3H_0^3 \Delta(a)^{0.35}} \int dz \frac{ R_*(z) }{(1+z)^{1.65} E(\Omega_M,\Omega_{\Lambda},z) } \nonumber \\
&& \Theta(f-f_{min}(a,z)) \Theta(f_{max}(a,z)-f),
\end{eqnarray}
which is identical to Equation \ref{omega_zint} that was obtained by evaluating the $z$-integral instead of the $m$-integral.

We note that the chemical evolution model that was used to determine the star formation rate \cite{vangioni} uses a value of $\alpha$ which varies between 0.08 and 0.30
depending on the progenitor mass and metallicity \cite{ww95}. However, we have confirmed that using the larger value of $\alpha=0.5$ in the chemical evolution model does not have a significant impact on the best fit for the star formation rate. In particular, increasing $\alpha$ reduces the amount of metals released in the core collapse events by 10-20\%, but the effect is well within the uncertainty on metallicity measurements.

\begin{figure}
\includegraphics[width=3.2in]{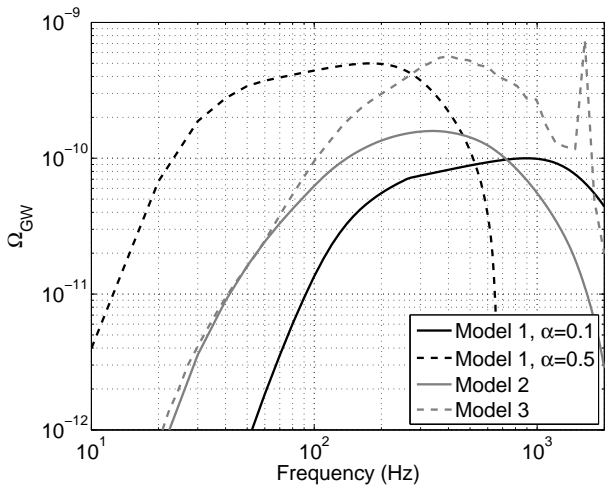}
  \caption{Comparison of Models 1, 2, and 3, for $\epsilon = 10^{-5}$ and $a = 0.2$.    }
  \label{model1vs2}
\end{figure}

Hence, Model 2 is characterized by 2 parameters, $\epsilon$ and $a$, which have similar effects on the GW energy spectrum as in Model 1. The third free parameter in Model 1, $\alpha$, is now modelled as a function of metallicity and redshift based on the population synthesis simulations. Figure \ref{model1vs2} shows a Model 2 spectrum in comparison with Model 1 spectra obtained for the same values of $\epsilon$ and $a$ and for the largest and smallest values of $\alpha$. Model 2 spectrum can be seen as an effective average of the Model 1 spectra.

\begin{figure}
\includegraphics[width=3.2in]{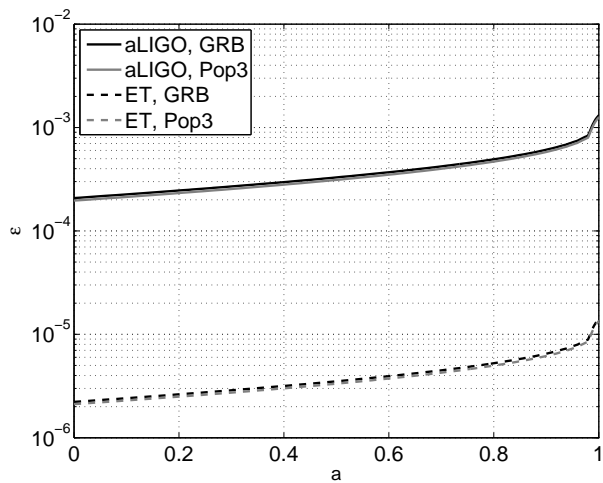}
  \caption{95\% confidence sensitivity contours of the Advanced LIGO and Einstein Telescope and for two models of star formation rate are shown in the $\epsilon-a$ plane for the Model 2.}
  \label{ccbh_a_metal}
\end{figure}

Figure \ref{ccbh_a_metal} shows the 95\% confidence contours for the Model 2 in the $\epsilon-a$ plane for the Advanced LIGO and Einstein Telescope detectors. Similarly to Model 1, the two different star formation models yield nearly identical results, and the expected ET sensitivity is about $100\times$ better than that for Advanced LIGO. The contours are also about $2\times$ lower in $\epsilon$ than for Model 1, which is a consequence of the fact that $\alpha(Z)$ is distributed differently in Models 1 and 2.

\section{Model 3}
We now treat the dependence of $\alpha$, $m_{min}$, and $m_{max}$ on metallicity more carefully, by using the results of {\tt StarTrack} \citep{Belczynski2002,Belczynski2008}, a sophisticated population synthesis code able to generate realistic populations
of single and binary compact objects (neutron stars and black holes).
The code is based on revised formulas from \cite{Hurley2000}; updated with
new wind mass loss prescriptions, calibrated tidal interactions, physical
estimation of donor's binding energy ($\lambda$) and convection driven,
neutrino enhanced supernova engines. A full description of these updates is
given in \cite{Dominik2012}. The two most recent updates take into account
measurements of initial parameter distributions for massive O stars
\citep{Sana2012} as well as a correction of a technical bug that has limited
the formation of BH-BH binaries for high metallicity (e.g., $Z=0.02$).

\begin{figure}
\includegraphics[width=3.2in]{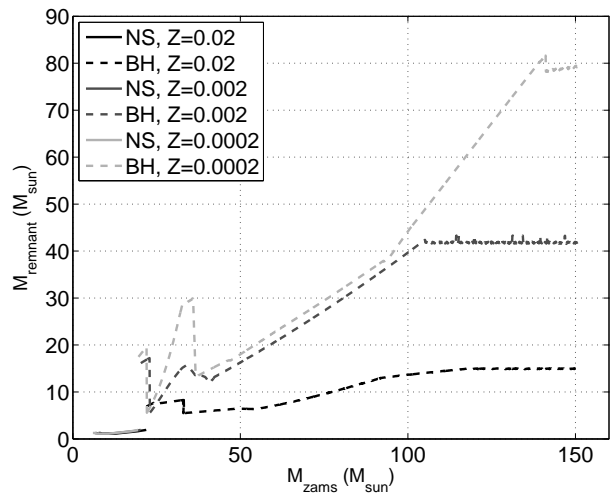}
  \caption{Final mass of a remnant as a function of its progenitor's mass and metallicity for single stellar evolution as predicted by our population synthesis model. Note that the maximum BH mass increases rapidly with decreasing metallicity. Maximum BH mass for solar metallicity is only $15 M_{\odot}$, while for 1\% solar metallicity it is $81M_{\odot}$. Our model also reproduces the observed mass gap between NSs and BHs. In our predictions there is no compact objects in mass range $1.8-5.5 M_{\odot}$ ($Z_{\odot}$), $1.9-5.6M_{\odot}$ (0.1 $Z_{\odot}$), and $1.8-6.7 M_{\odot}$ (0.01 $Z_{\odot}$).}
  \label{StarTrack_mass}
\end{figure}

We evolve single
stars until the formation of a compact object. Simulations are done for a
dense grid of stellar metallicities: $Z=0.0001$--$0.03$ with step of
$\Delta Z =0.0001$ (solar composition is $Z=Z_\odot \approx 0.02$). Two
major factors shape the initial (Zero Age Main Sequence) -- final (compact
object) mass relation: wind mass loss and core collapse/supernova compact
object formation. For wind mass loss we use O/B type winds from \cite{Vink2001}
and for other evolutionary stages (e.g., LBV winds) formulae as calibrated in
\cite{belczynski}. We adopt the set of models presented by \citep{Fryer2012}
with the rapid core collapse/supernova mechanism. The explosion occurs within
the first $0.1-0.2$s driven by a convection and neutrino enhanced engine. This
engine reproduces \citep{Belczynski2012} the mass gap observed in Galactic X-ray
binaries \citep{Ozel, Bailyn}. The typical initial-final mass
relations are presented in Figure~\ref{StarTrack_mass}.

Pair-instability supernovae (PISNe) may disrupt the most massive stars
without leaving behind a compact object. The original simulations indicate
that stars in the initial mass range $M_{\rm zams}=150-300 M_{\odot}$ may be
subject to PISNe \cite{Fryer2001} (here, the subscript "zams" stands for the "zero-age mass sequence"). These initial estimates were translated
into the final CO core mass (as this can be used for various metallicities
and wind mass loss rates). The range for PISNe to disrupt stars without
black hole formation is found approximately in range $M_{\rm CO}=60-130 M_{\odot}$
\cite{Yusof2013}. For our assumed extent of IMF (up to 150 $M_{\odot}$) and the range
of employed metallicities we are not in the regime of pair-instability supernovae
(PISNe).
However, recent discovery of stars as massive as $\sim 200-300 M_{\odot}$ in the
R136 star cluster in LMC \cite{Crowther2010} may possibly indicate that IMF
extends beyond PISNe regime, implying that stars more massive than $300 M_{\odot}$ may exist and
form massive BHs ($\gtrsim 100 M_{\odot}$) \cite{Belczynski2014}.


The evolutionary code {\tt StarTrack} provides $\alpha(Z,m)$ for a grid of metallicity in the interval $0.0001-0.03$, and progenitor mass between $7-150$ M$_{\odot}$. It also gives $m_{\min}(Z)$ and $m_{\max}(Z)$, the minimal and the maximal masses to form a black hole for a given metallicity.
Using these data, we can simulate a population of massive stars that undergo core-collapse to black holes. Our Monte-Carlo procedure is described below.

We fix the spin parameter $a$ and the efficiency $\varepsilon$, and we proceeded as follows for $N_{\rm MC}=10^6$ sources :
\begin{itemize}
\item We draw the redshift $z$ from a probability distribution constructed by normalizing the rate $R_z(z)$ in the interval $0-20$.
\begin{equation}
p_z(z) = \frac{R_z(z)}{\dot{N}}
\end{equation}
where $\dot{N}=\int_0^{20} R_z(z) dz$
and calculate the luminosity distance $d_L(z) = (1+z)r(z)$.

\item We calculate the average metallicity $\bar{Z}$ at redshift $z$ and draw the metallicity from a uniform distribution in the interval $[\bar{Z}/3-3\bar{Z}]$.


\item By interpolating in the {\tt StarTrack} data, we calculate $m_{\min}(Z)$ and $m_{\max}(Z)$, as well as the corresponding mass fraction $\lambda_{BH}(Z)$ (c.f. Eq. \ref{eqrv}). For all metallicities, the maximal mass is assumed to be 150 M$_{\odot}$ while the minimal mass increases from 20 M$_{\odot}$ for $Z=0.0001$ to 36 M$_{\odot}$ for $Z=0.03$.
We also draw the mass of the progenitor from a distribution constructed from the initial mass function:
\begin{equation}
p_m(m,Z)= \frac{\phi(m)}{ \int_{m_{\min}(Z)}^{m_{\max}(Z)}\phi(m) dm}
\end{equation}
We used two different models: the Saltpeter IMF introduced in the previous sections and a three component broken
power-law IMF with slope of $-1.3$ for initial mass
$M_{\rm zams}=0.08 - 0.5M_{\odot}$, $-2.2$ for $M_{\rm zams}=0.5 - 1 M_{\odot}$, and $-2.7$
for $M_{\rm zams}=1 - 150 M_{\odot}$ \cite{Kroupa2003}.

\item  We deduce $\alpha(Z,m)$ and then the mass of the BH $M_{BH}(m,Z)=\alpha(Z,m) m$, by interpolating in the {\tt StarTrack} data.

\item We calculate the emission frequency $\nu_*(m,a,\alpha(Z,m))$ and the observed frequency $f=(1+z) \nu_*$.
\end{itemize}

The sources are stored into frequency bins of length $\Delta f = 10$ Hz, with central frequencies $f_i=10i +5$ Hz for i=1...500.
The energy density at the frequency $f_i$ is given by the discrete sum of the individual contribution of the $N_i$ sources in the $i^{th}$ bin:

\begin{equation}
\Omega_{\rm GW} (f_i)= \frac{8 \pi G}{3 H_0^2 c} \frac{\dot{N}}{\Delta f N_{\rm MC}} f_i  \sum_{k=0}^{N_i} \lambda_{BH}(Z) \frac{\varepsilon  M_{BH}(Z) (1+z)} {4 \pi d_{L}(z)^2 }
\end{equation}

The spectrum for $a=0.2$ and $\varepsilon=10^{-5}$ is shown in Figure \ref{model1vs2} and compared to Models 1 and 2. The fluctuation at high frequencies is due to the small number sources in the simulation at such high frequencies. The above procedure is repeated for a grid of points in the $\epsilon-a$ parameter space, producing $\Omega_{\rm GW}(f)$ for each point in this grid. The gravitational-wave spectra are then compared to the projected sensitivities of the second and third generation detectors, similarly to Models 1 and 2, using the likelihood formalism (c.f. Equation \ref{likelihood_eq}).

Finally, to assess the importance of possible black hole progenitors of masses beyond our assumed upper limit of $150 M_{\odot}$, we repeat the above calculations extending the IMF up to $300 M_{\odot}$ and $1000 M_{\odot}$ under the assumption that all stars above $300 M_{\odot}$ produce $100 M_{\odot}$ black holes.

\begin{figure*}[!t]
   \begin{tabular}{ccc}
\includegraphics[width=2.5in]{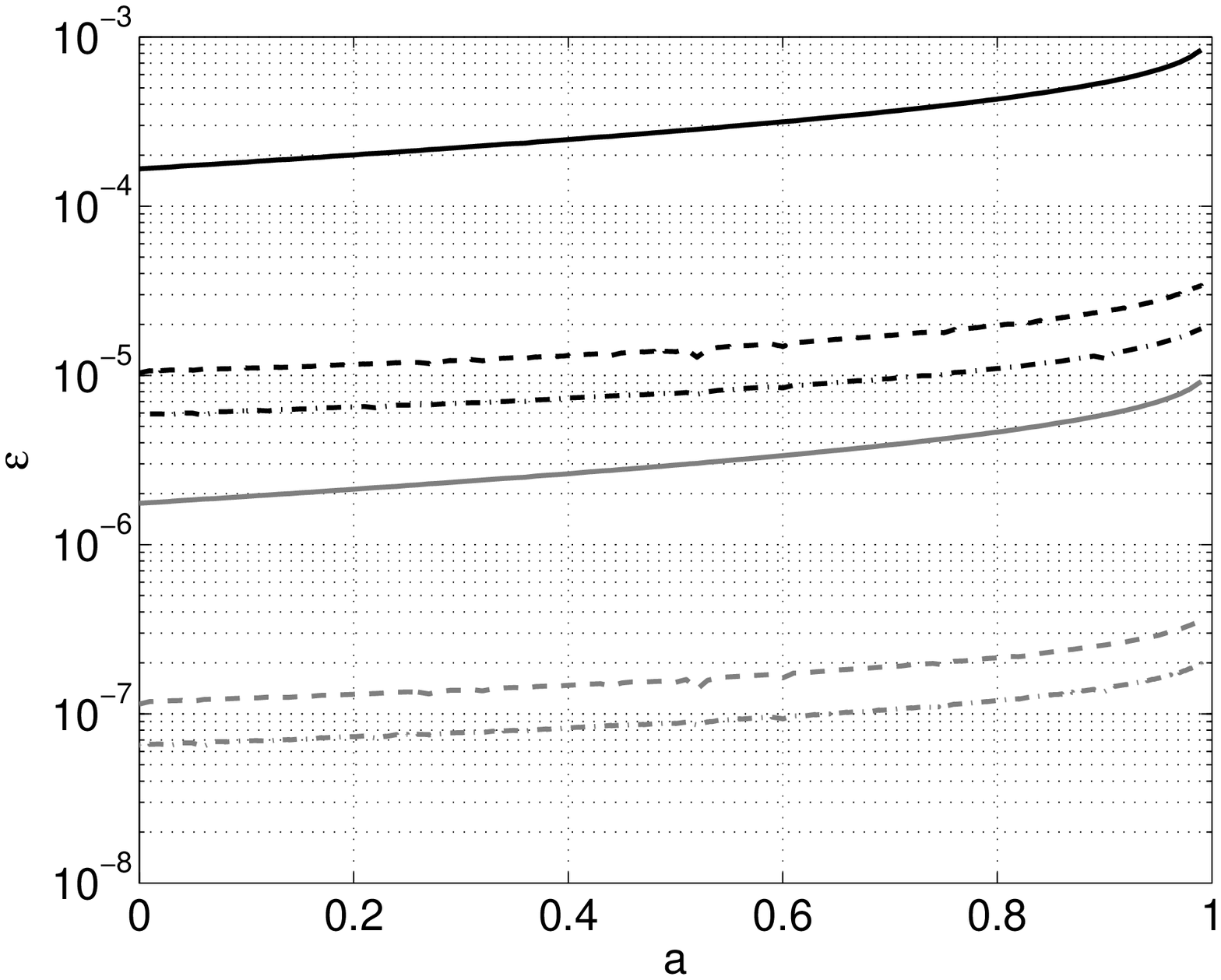} &
\includegraphics[width=1.4in]{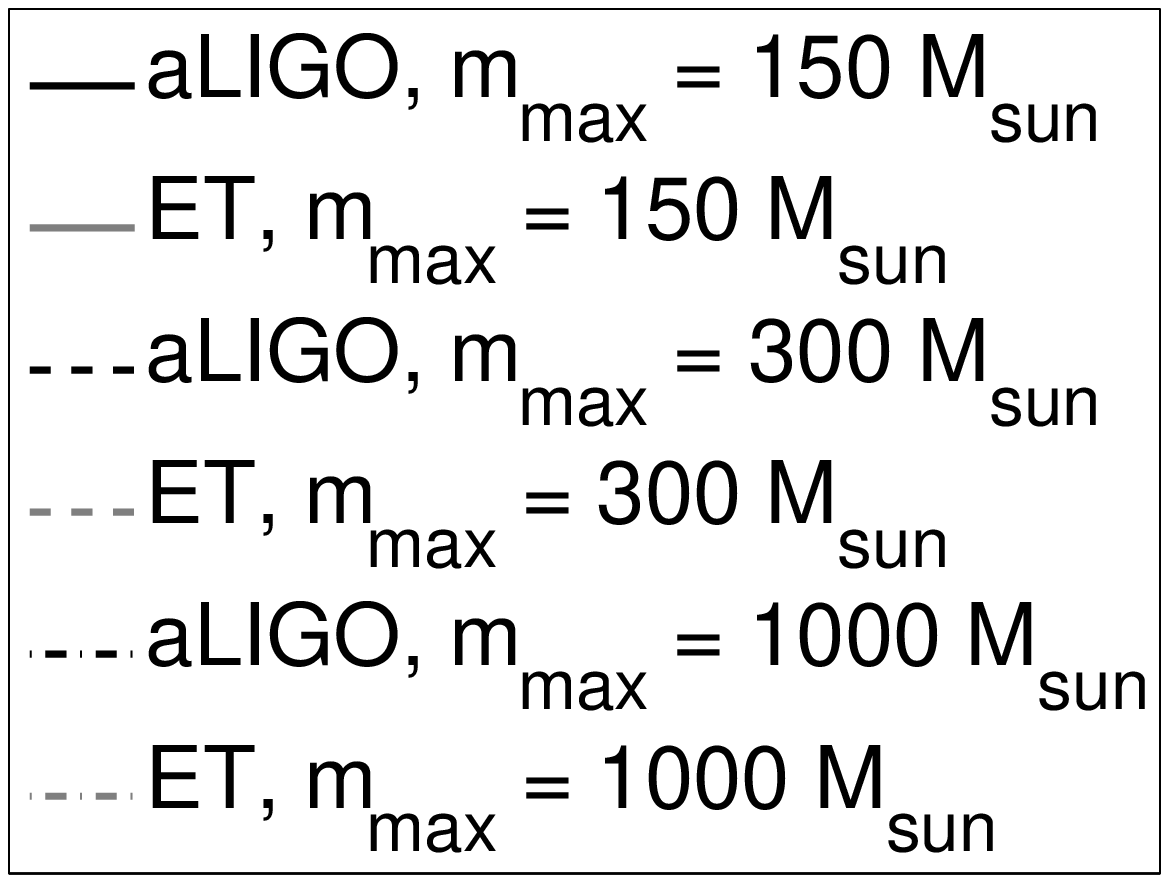} &
\includegraphics[width=2.5in]{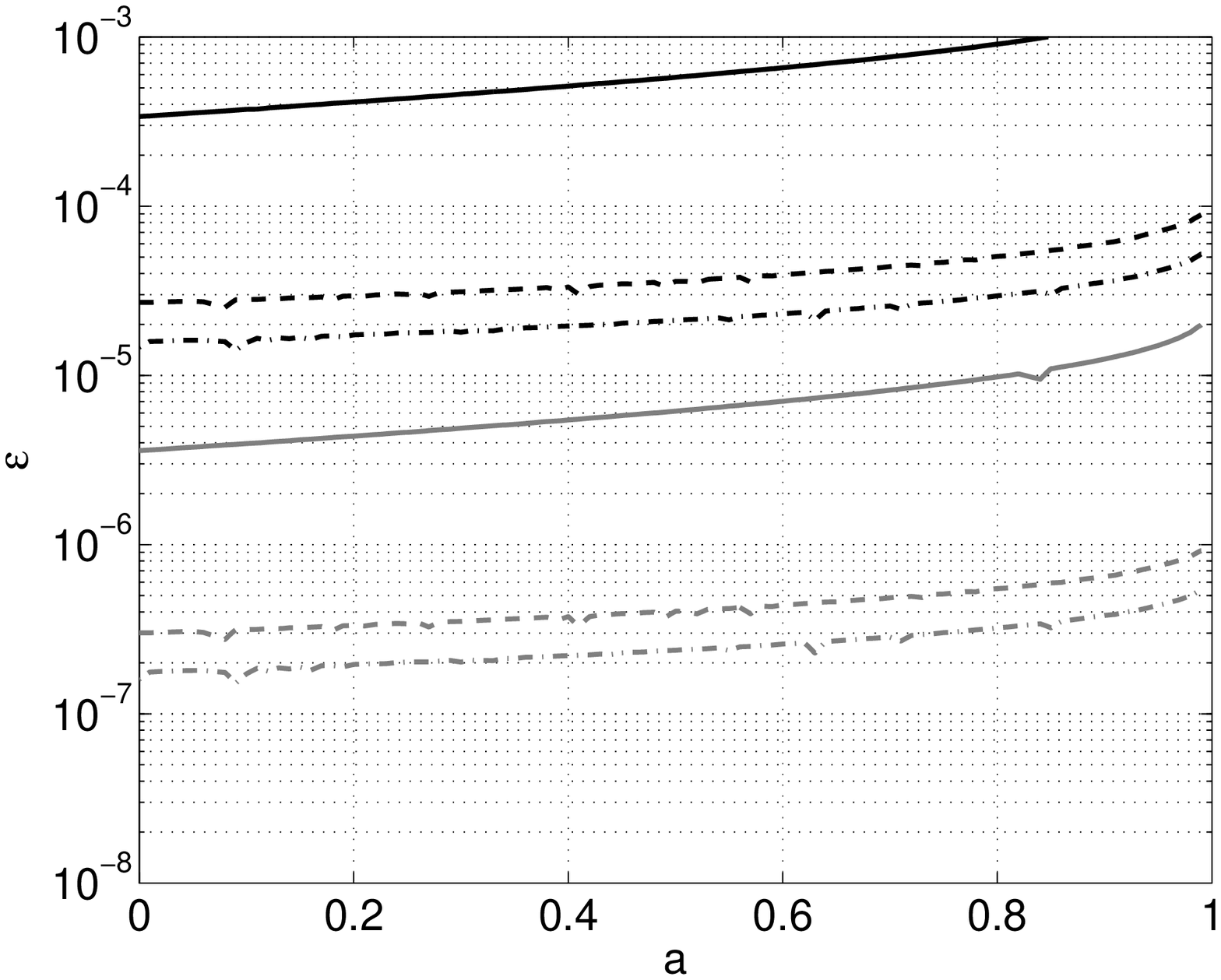}\\
  \end{tabular}
  \caption{95\% confidence sensitivity contours for the Advanced LIGO (black) and Einstein Telescope (gray) and for three values of $m_{\rm max}=150 M_{\odot}$ (solid), $300 M_{\odot}$ (dashed), and $1000 M_{\odot}$ (dot-dashed) are shown in the $\epsilon-a$ plane for the Model 3, assuming the Salpeter IMF (left plot) and the Koupra IMF (right plot).      }
  \label{model3_epsilona}
\end{figure*}

The 95\% sensitivities of future detectors are shown in Figure \ref{model3_epsilona}, in the $\epsilon-a$ plane. Note that the sensitivity contours are about $2\times$ lower in $\epsilon$ relative to Model 2, which is the result of the more careful treatment of the dependence of $\alpha$ on the metallicity, including the PISNe mass region. Changing the value of maximum progenitor mass again has a significant impact on detectability of this model, as already observed in Model 1. The two IMF models considered in this calculation agree to within a factor of about 2. We have verified that the two different SFR models considered in this paper yield nearly identical results, as already observed in Models 1 and 2.

Finally, as noted in the case of Model 1, the realistic values of $\epsilon \sim 10^{-7}-10^{-6}$ \cite{baiotti} may be reached by Einstein Telescope, especially in the case if massive ($>300 M_{\odot}$) black hole progenitors exist, as suggested by recent observations. The second generation detectors may probe the more exotic scenarios associated with $\epsilon \sim 0.01$ \cite{kobayashi}.

\section{Conclusions}
In this paper we have studied the stochastic gravitational wave background generated by stellar core collapse events occurring throughout the universe. While the core collapse process is likely associated with several mechanisms for GW production, we focused on the most tractable one, namely the ringdown of the newly formed black hole, following the collapse itself. While this mechanism has been studied in the past, our purpose here is to revisit it in more detail, and study the detectability of the corresponding background by the upcoming second and third generation gravitational-wave detectors. We considered three variations of the model. Our Model 1 assumes that the dominant gravitational-wave mode during the ringdown is $l=2$, and it integrates this signal across all core collapse events that yield black holes. For this purpose, we used the most recent models of star formation rate that are consistent with the metallicity observations as well as with the CMB-based constraints on the reionization redshift and the optical depth parameters. We scan the parameter space of the model and determine the part of the parameter space that is accessible to future detectors. In our Model 2, we attempt to include the effect of metallicity (which evolves with redshift) on the mass of the newly produced black hole. In this model we treat this effect in an average sense, allowing us to compute the gravitational-wave spectrum analytically. In Model 3, we go a step further and instead of an analytic calculation, we perform Monte Carlo simulations, drawing the stars from the appropriate redshift and mass distributions determined by the StarTrack simulations.

Remarkably, the three models agree with each other to within a factor of $\sim 2$, indicating their robustness. However, the required efficiency of gravitational-wave production by the ringdown is relatively high. In particular, it is unlikely that the second generation detectors will observe this background since the required efficiency of gravitational-wave production is $\epsilon \sim 10^{-4}$ or higher, which is not supported by simulations. The third generation detectors, however, would require $\epsilon \sim 10^{-6}$ or higher, which is more realistic. We emphasize, however, that we have considered only one of the several mechanisms for gravitational-wave production in core collapse processes. Future studies should attempt to repeat similar studies for other mechanisms---while other mechanisms are more difficult to model (due to the complexity of the relevant processes), they are also likely to produce a stronger gravitational-wave background. Furthermore, since most massive stars reside in binaries, we expect that inclusion of binary systems in this calculation may have a significant impact on the gravitational-wave background estimate. We plan to include this step in a follow-up study.


\section*{Acknowledgments}
The work K.A.O. was supported in part by DOE grant DE-SC0011842  at the University of Minnesota. The work of K.C. and V.M. was supported in part by NSF grant PHY-1204944 at the University of Minnesota. KB and WG evolutionary modeling was partially sponsored by the NCN grant Sonata Bis 2  (DEC-2012/07/E/ST9/01360) and Polish Science Foundation (FNP) "Master 2013" Subsidy. The work  of EV has been carried out at the ILP LABEX (under reference ANR-10-LABX-63) supported by French state funds managed by the ANR within the Investissements d’Avenir programme under reference ANR-11-IDEX-0004-02, and was also sponsored by the French Agence Nationale pour la Recherche (ANR) via the grant VACOUL (ANR-2010-Blan-0510-01). LIGO-P1500077.
\bibliography{ccbh}

\begin{thebibliography}{114}
\expandafter\ifx\csname natexlab\endcsname\relax\def\natexlab#1{#1}\fi
\expandafter\ifx\csname bibnamefont\endcsname\relax
  \def\bibnamefont#1{#1}\fi
\expandafter\ifx\csname bibfnamefont\endcsname\relax
  \def\bibfnamefont#1{#1}\fi
\expandafter\ifx\csname citenamefont\endcsname\relax
  \def\citenamefont#1{#1}\fi
\expandafter\ifx\csname url\endcsname\relax
  \def\url#1{\texttt{#1}}\fi
\expandafter\ifx\csname urlprefix\endcsname\relax\def\urlprefix{URL }\fi
\providecommand{\bibinfo}[2]{#2}
\providecommand{\eprint}[2][]{\url{#2}}

\bibitem[{\citenamefont{Grishchuk}(1975)}]{grishchuk}
\bibinfo{author}{\bibfnamefont{L.~P.} \bibnamefont{Grishchuk}},
  \bibinfo{journal}{Sov. Phys. JETP} \textbf{\bibinfo{volume}{40}},
  \bibinfo{pages}{409} (\bibinfo{year}{1975}).

\bibitem[{\citenamefont{Starobinskii}(1979)}]{starob}
\bibinfo{author}{\bibfnamefont{A.~A.} \bibnamefont{Starobinskii}},
  \bibinfo{journal}{JETP Lett.} \textbf{\bibinfo{volume}{30}},
  \bibinfo{pages}{682} (\bibinfo{year}{1979}).

\bibitem[{\citenamefont{Easther and Lim}(2006)}]{eastherlim}
\bibinfo{author}{\bibfnamefont{R.}~\bibnamefont{Easther}} \bibnamefont{and}
  \bibinfo{author}{\bibfnamefont{E.~A.} \bibnamefont{Lim}},
  \bibinfo{journal}{JCAP} \textbf{\bibinfo{volume}{0604}}, \bibinfo{pages}{010}
  (\bibinfo{year}{2006}).

\bibitem[{\citenamefont{Barnaby et~al.}(2012)\citenamefont{Barnaby, Pajer, and
  Peloso}}]{peloso}
\bibinfo{author}{\bibfnamefont{N.}~\bibnamefont{Barnaby}},
  \bibinfo{author}{\bibfnamefont{E.}~\bibnamefont{Pajer}}, \bibnamefont{and}
  \bibinfo{author}{\bibfnamefont{M.}~\bibnamefont{Peloso}},
  \bibinfo{journal}{Phys. Rev. D} \textbf{\bibinfo{volume}{85}},
  \bibinfo{pages}{023525} (\bibinfo{year}{2012}).

\bibitem[{\citenamefont{Caldwell and Allen}(1992)}]{caldwellallen}
\bibinfo{author}{\bibfnamefont{R.~R.} \bibnamefont{Caldwell}} \bibnamefont{and}
  \bibinfo{author}{\bibfnamefont{B.}~\bibnamefont{Allen}},
  \bibinfo{journal}{Phys. Rev. D} \textbf{\bibinfo{volume}{45}},
  \bibinfo{pages}{3447} (\bibinfo{year}{1992}).

\bibitem[{\citenamefont{Damour and Vilenkin}(2000)}]{DV1}
\bibinfo{author}{\bibfnamefont{T.}~\bibnamefont{Damour}} \bibnamefont{and}
  \bibinfo{author}{\bibfnamefont{A.}~\bibnamefont{Vilenkin}},
  \bibinfo{journal}{Phys. Rev. Lett.} \textbf{\bibinfo{volume}{85}},
  \bibinfo{pages}{3761} (\bibinfo{year}{2000}).

\bibitem[{\citenamefont{Damour and Vilenkin}(2005)}]{DV2}
\bibinfo{author}{\bibfnamefont{T.}~\bibnamefont{Damour}} \bibnamefont{and}
  \bibinfo{author}{\bibfnamefont{A.}~\bibnamefont{Vilenkin}},
  \bibinfo{journal}{Phys. Rev. D} \textbf{\bibinfo{volume}{71}},
  \bibinfo{pages}{063510} (\bibinfo{year}{2005}).

\bibitem[{\citenamefont{Siemens et~al.}(2007)\citenamefont{Siemens, Mandic, and
  Creighton}}]{cosmstrpaper}
\bibinfo{author}{\bibfnamefont{X.}~\bibnamefont{Siemens}},
  \bibinfo{author}{\bibfnamefont{V.}~\bibnamefont{Mandic}}, \bibnamefont{and}
  \bibinfo{author}{\bibfnamefont{J.}~\bibnamefont{Creighton}},
  \bibinfo{journal}{Phys. Rev. Lett.} \textbf{\bibinfo{volume}{98}},
  \bibinfo{pages}{111101} (\bibinfo{year}{2007}).

\bibitem[{\citenamefont{Olmez et~al.}(2010)\citenamefont{Olmez, Mandic, and
  Siemens}}]{olmez1}
\bibinfo{author}{\bibfnamefont{S.}~\bibnamefont{Olmez}},
  \bibinfo{author}{\bibfnamefont{V.}~\bibnamefont{Mandic}}, \bibnamefont{and}
  \bibinfo{author}{\bibfnamefont{X.}~\bibnamefont{Siemens}},
  \bibinfo{journal}{Phys. Rev. D} \textbf{\bibinfo{volume}{81}},
  \bibinfo{pages}{104028} (\bibinfo{year}{2010}).

\bibitem[{\citenamefont{Olmez et~al.}(2012)\citenamefont{Olmez, Mandic, and
  Siemens}}]{olmez2}
\bibinfo{author}{\bibfnamefont{S.}~\bibnamefont{Olmez}},
  \bibinfo{author}{\bibfnamefont{V.}~\bibnamefont{Mandic}}, \bibnamefont{and}
  \bibinfo{author}{\bibfnamefont{X.}~\bibnamefont{Siemens}},
  \bibinfo{journal}{J. Cosm. Astrop. Phys.} \textbf{\bibinfo{volume}{07}},
  \bibinfo{pages}{009} (\bibinfo{year}{2012}).

\bibitem[{\citenamefont{Gasperini and Veneziano}(1993)}]{PBB1}
\bibinfo{author}{\bibfnamefont{M.}~\bibnamefont{Gasperini}} \bibnamefont{and}
  \bibinfo{author}{\bibfnamefont{G.}~\bibnamefont{Veneziano}},
  \bibinfo{journal}{Astropart. Phys.} \textbf{\bibinfo{volume}{1}},
  \bibinfo{pages}{317} (\bibinfo{year}{1993}).

\bibitem[{\citenamefont{Mandic and Buonanno}(2006)}]{PBBpaper}
\bibinfo{author}{\bibfnamefont{V.}~\bibnamefont{Mandic}} \bibnamefont{and}
  \bibinfo{author}{\bibfnamefont{A.}~\bibnamefont{Buonanno}},
  \bibinfo{journal}{Phys. Rev. D} \textbf{\bibinfo{volume}{73}},
  \bibinfo{pages}{063008} (\bibinfo{year}{2006}).

\bibitem[{\citenamefont{Regimbau}(2011)}]{regimbau}
\bibinfo{author}{\bibfnamefont{T.}~\bibnamefont{Regimbau}},
  \bibinfo{journal}{Res. Astr. Astrop.} \textbf{\bibinfo{volume}{11}},
  \bibinfo{pages}{369} (\bibinfo{year}{2011}).

\bibitem[{\citenamefont{Phinney}(1991)}]{phinney}
\bibinfo{author}{\bibfnamefont{E.~S.} \bibnamefont{Phinney}},
  \bibinfo{journal}{Astrophys. J. Lett.} \textbf{\bibinfo{volume}{380}},
  \bibinfo{pages}{L17} (\bibinfo{year}{1991}).

\bibitem[{\citenamefont{Kosenko and Postnov}(1998)}]{kosenko}
\bibinfo{author}{\bibfnamefont{D.~I.} \bibnamefont{Kosenko}} \bibnamefont{and}
  \bibinfo{author}{\bibfnamefont{K.~A.} \bibnamefont{Postnov}},
  \bibinfo{journal}{Astron. \& Astrop.} \textbf{\bibinfo{volume}{336}},
  \bibinfo{pages}{786} (\bibinfo{year}{1998}).

\bibitem[{\citenamefont{{Regimbau} and {de Freitas
  Pacheco}}(2006{\natexlab{a}})}]{2006ApJ...642..455R}
\bibinfo{author}{\bibfnamefont{T.}~\bibnamefont{{Regimbau}}} \bibnamefont{and}
  \bibinfo{author}{\bibfnamefont{J.~A.} \bibnamefont{{de Freitas Pacheco}}},
  \bibinfo{journal}{\apj} \textbf{\bibinfo{volume}{642}}, \bibinfo{pages}{455}
  (\bibinfo{year}{2006}{\natexlab{a}}), \eprint{gr-qc/0512008}.

\bibitem[{\citenamefont{Zhu et~al.}(2011)\citenamefont{Zhu, Howell, Regimbau,
  Blair, and Zhu}}]{zhu}
\bibinfo{author}{\bibfnamefont{X.-J.} \bibnamefont{Zhu}},
  \bibinfo{author}{\bibfnamefont{E.}~\bibnamefont{Howell}},
  \bibinfo{author}{\bibfnamefont{T.}~\bibnamefont{Regimbau}},
  \bibinfo{author}{\bibfnamefont{D.}~\bibnamefont{Blair}}, \bibnamefont{and}
  \bibinfo{author}{\bibfnamefont{Z.-H.} \bibnamefont{Zhu}},
  \bibinfo{journal}{Astrophys. J.} \textbf{\bibinfo{volume}{739}},
  \bibinfo{pages}{86} (\bibinfo{year}{2011}).

\bibitem[{\citenamefont{{Rosado}}(2011)}]{2011PhRvD..84h4004R}
\bibinfo{author}{\bibfnamefont{P.~A.} \bibnamefont{{Rosado}}},
  \bibinfo{journal}{\prd} \textbf{\bibinfo{volume}{84}}, \bibinfo{eid}{084004}
  (\bibinfo{year}{2011}), \eprint{1106.5795}.

\bibitem[{\citenamefont{{Marassi}
  et~al.}(2011{\natexlab{a}})\citenamefont{{Marassi}, {Schneider}, {Corvino},
  {Ferrari}, and {Zwart}}}]{2011PhRvD..84l4037M}
\bibinfo{author}{\bibfnamefont{S.}~\bibnamefont{{Marassi}}},
  \bibinfo{author}{\bibfnamefont{R.}~\bibnamefont{{Schneider}}},
  \bibinfo{author}{\bibfnamefont{G.}~\bibnamefont{{Corvino}}},
  \bibinfo{author}{\bibfnamefont{V.}~\bibnamefont{{Ferrari}}},
  \bibnamefont{and} \bibinfo{author}{\bibfnamefont{S.~P.}
  \bibnamefont{{Zwart}}}, \bibinfo{journal}{\prd}
  \textbf{\bibinfo{volume}{84}}, \bibinfo{eid}{124037}
  (\bibinfo{year}{2011}{\natexlab{a}}), \eprint{1111.6125}.

\bibitem[{\citenamefont{Wu et~al.}(2012)\citenamefont{Wu, Mandic, and
  Regimbau}}]{StochCBC}
\bibinfo{author}{\bibfnamefont{C.}~\bibnamefont{Wu}},
  \bibinfo{author}{\bibfnamefont{V.}~\bibnamefont{Mandic}}, \bibnamefont{and}
  \bibinfo{author}{\bibfnamefont{T.}~\bibnamefont{Regimbau}},
  \bibinfo{journal}{Phys. Rev. D} \textbf{\bibinfo{volume}{85}},
  \bibinfo{pages}{104024} (\bibinfo{year}{2012}).

\bibitem[{\citenamefont{{Zhu} et~al.}(2013)\citenamefont{{Zhu}, {Howell},
  {Blair}, and {Zhu}}}]{2013MNRAS.431..882Z}
\bibinfo{author}{\bibfnamefont{X.-J.} \bibnamefont{{Zhu}}},
  \bibinfo{author}{\bibfnamefont{E.~J.} \bibnamefont{{Howell}}},
  \bibinfo{author}{\bibfnamefont{D.~G.} \bibnamefont{{Blair}}},
  \bibnamefont{and} \bibinfo{author}{\bibfnamefont{Z.-H.} \bibnamefont{{Zhu}}},
  \bibinfo{journal}{Monthly Notices of the Royal astronomical Society}
  \textbf{\bibinfo{volume}{431}}, \bibinfo{pages}{882} (\bibinfo{year}{2013}),
  \eprint{1209.0595}.

\bibitem[{\citenamefont{{Evangelista} and {de
  Araujo}}(2013)}]{2013MPLA...2850174E}
\bibinfo{author}{\bibfnamefont{E.~F.~D.} \bibnamefont{{Evangelista}}}
  \bibnamefont{and} \bibinfo{author}{\bibfnamefont{J.~C.~N.} \bibnamefont{{de
  Araujo}}}, \bibinfo{journal}{Modern Physics Letters A}
  \textbf{\bibinfo{volume}{28}}, \bibinfo{eid}{1350174} (\bibinfo{year}{2013}),
  \eprint{1504.04300}.

\bibitem[{\citenamefont{{Kowalska-Leszczynska}
  et~al.}(2015)\citenamefont{{Kowalska-Leszczynska}, {Regimbau}, {Bulik},
  {Dominik}, and {Belczynski}}}]{2015A&A...574A..58K}
\bibinfo{author}{\bibfnamefont{I.}~\bibnamefont{{Kowalska-Leszczynska}}},
  \bibinfo{author}{\bibfnamefont{T.}~\bibnamefont{{Regimbau}}},
  \bibinfo{author}{\bibfnamefont{T.}~\bibnamefont{{Bulik}}},
  \bibinfo{author}{\bibfnamefont{M.}~\bibnamefont{{Dominik}}},
  \bibnamefont{and}
  \bibinfo{author}{\bibfnamefont{K.}~\bibnamefont{{Belczynski}}},
  \bibinfo{journal}{A\&A} \textbf{\bibinfo{volume}{574}}, \bibinfo{eid}{A58}
  (\bibinfo{year}{2015}), \eprint{1205.4621}.

\bibitem[{\citenamefont{{Evangelista} and {de
  Araujo}}(2015)}]{2015MNRAS.449.2700E}
\bibinfo{author}{\bibfnamefont{E.~F.~D.} \bibnamefont{{Evangelista}}}
  \bibnamefont{and} \bibinfo{author}{\bibfnamefont{J.~C.~N.} \bibnamefont{{de
  Araujo}}}, \bibinfo{journal}{Mon. Not. Roy. Astron. Soc.}
  \textbf{\bibinfo{volume}{449}}, \bibinfo{pages}{2700} (\bibinfo{year}{2015}),
  \eprint{1504.02700}.

\bibitem[{\citenamefont{Regimbau and de~Freitas~Pacheco}(2001)}]{RegPac}
\bibinfo{author}{\bibfnamefont{T.}~\bibnamefont{Regimbau}} \bibnamefont{and}
  \bibinfo{author}{\bibfnamefont{J.~A.} \bibnamefont{de~Freitas~Pacheco}},
  \bibinfo{journal}{Astron. and Astrophys.} \textbf{\bibinfo{volume}{376}},
  \bibinfo{pages}{381} (\bibinfo{year}{2001}).

\bibitem[{\citenamefont{Owen et~al.}(1998)\citenamefont{Owen, Lindblom, Cutler,
  Schutz, Vecchio, and Andersson}}]{owen}
\bibinfo{author}{\bibfnamefont{B.~J.} \bibnamefont{Owen}},
  \bibinfo{author}{\bibfnamefont{L.}~\bibnamefont{Lindblom}},
  \bibinfo{author}{\bibfnamefont{C.}~\bibnamefont{Cutler}},
  \bibinfo{author}{\bibfnamefont{B.~F.} \bibnamefont{Schutz}},
  \bibinfo{author}{\bibfnamefont{A.}~\bibnamefont{Vecchio}}, \bibnamefont{and}
  \bibinfo{author}{\bibfnamefont{N.}~\bibnamefont{Andersson}},
  \bibinfo{journal}{Phys. Rev. D} \textbf{\bibinfo{volume}{58}},
  \bibinfo{pages}{084020} (\bibinfo{year}{1998}).

\bibitem[{\citenamefont{{Ferrari}
  et~al.}(1999{\natexlab{a}})\citenamefont{{Ferrari}, {Matarrese}, and
  {Schneider}}}]{1999MNRAS.303..258F}
\bibinfo{author}{\bibfnamefont{V.}~\bibnamefont{{Ferrari}}},
  \bibinfo{author}{\bibfnamefont{S.}~\bibnamefont{{Matarrese}}},
  \bibnamefont{and}
  \bibinfo{author}{\bibfnamefont{R.}~\bibnamefont{{Schneider}}},
  \bibinfo{journal}{Monthly Notices of the Royal astronomical Society}
  \textbf{\bibinfo{volume}{303}}, \bibinfo{pages}{258}
  (\bibinfo{year}{1999}{\natexlab{a}}), \eprint{astro-ph/9806357}.

\bibitem[{\citenamefont{Chandrasekhar}(1969)}]{barmodes1}
\bibinfo{author}{\bibfnamefont{S.}~\bibnamefont{Chandrasekhar}},
  \emph{\bibinfo{title}{"Ellipsoidal Figures of Equilibrium"}}
  (\bibinfo{publisher}{New Haven, Yale Univ. Press}, \bibinfo{year}{1969}).

\bibitem[{\citenamefont{Houser et~al.}(1994)\citenamefont{Houser, Centrella,
  and Smith}}]{barmodes2}
\bibinfo{author}{\bibfnamefont{J.~L.} \bibnamefont{Houser}},
  \bibinfo{author}{\bibfnamefont{J.~M.} \bibnamefont{Centrella}},
  \bibnamefont{and} \bibinfo{author}{\bibfnamefont{S.~C.} \bibnamefont{Smith}},
  \bibinfo{journal}{Phys. Rev. Lett.} \textbf{\bibinfo{volume}{72}},
  \bibinfo{pages}{1314} (\bibinfo{year}{1994}).

\bibitem[{\citenamefont{Lai and Shapiro}(1995)}]{barmodes3}
\bibinfo{author}{\bibfnamefont{D.}~\bibnamefont{Lai}} \bibnamefont{and}
  \bibinfo{author}{\bibfnamefont{S.~L.} \bibnamefont{Shapiro}},
  \bibinfo{journal}{Astrophys. J.} \textbf{\bibinfo{volume}{442}},
  \bibinfo{pages}{259} (\bibinfo{year}{1995}).

\bibitem[{\citenamefont{{Howell} et~al.}(2004)\citenamefont{{Howell}, {Coward},
  {Burman}, {Blair}, and {Gilmore}}}]{2004MNRAS.351.1237H}
\bibinfo{author}{\bibfnamefont{E.}~\bibnamefont{{Howell}}},
  \bibinfo{author}{\bibfnamefont{D.}~\bibnamefont{{Coward}}},
  \bibinfo{author}{\bibfnamefont{R.}~\bibnamefont{{Burman}}},
  \bibinfo{author}{\bibfnamefont{D.}~\bibnamefont{{Blair}}}, \bibnamefont{and}
  \bibinfo{author}{\bibfnamefont{J.}~\bibnamefont{{Gilmore}}},
  \bibinfo{journal}{Monthly Notices of the Royal astronomical Society}
  \textbf{\bibinfo{volume}{351}}, \bibinfo{pages}{1237} (\bibinfo{year}{2004}).

\bibitem[{\citenamefont{{Zhu} et~al.}(2011)\citenamefont{{Zhu}, {Fan}, and
  {Zhu}}}]{2011ApJ...729...59Z}
\bibinfo{author}{\bibfnamefont{X.-J.} \bibnamefont{{Zhu}}},
  \bibinfo{author}{\bibfnamefont{X.-L.} \bibnamefont{{Fan}}}, \bibnamefont{and}
  \bibinfo{author}{\bibfnamefont{Z.-H.} \bibnamefont{{Zhu}}},
  \bibinfo{journal}{\apj} \textbf{\bibinfo{volume}{729}}, \bibinfo{eid}{59}
  (\bibinfo{year}{2011}), \eprint{1102.2786}.

\bibitem[{\citenamefont{{Rosado}}(2012)}]{2012PhRvD..86j4007R}
\bibinfo{author}{\bibfnamefont{P.~A.} \bibnamefont{{Rosado}}},
  \bibinfo{journal}{\prd} \textbf{\bibinfo{volume}{86}}, \bibinfo{eid}{104007}
  (\bibinfo{year}{2012}), \eprint{1206.1330}.

\bibitem[{\citenamefont{{Lasky} et~al.}(2013)\citenamefont{{Lasky}, {Bennett},
  and {Melatos}}}]{2013PhRvD..87f3004L}
\bibinfo{author}{\bibfnamefont{P.~D.} \bibnamefont{{Lasky}}},
  \bibinfo{author}{\bibfnamefont{M.~F.} \bibnamefont{{Bennett}}},
  \bibnamefont{and}
  \bibinfo{author}{\bibfnamefont{A.}~\bibnamefont{{Melatos}}},
  \bibinfo{journal}{\prd} \textbf{\bibinfo{volume}{87}}, \bibinfo{eid}{063004}
  (\bibinfo{year}{2013}), \eprint{1302.6033}.

\bibitem[{\citenamefont{Cutler}(2002)}]{cutler}
\bibinfo{author}{\bibfnamefont{C.}~\bibnamefont{Cutler}},
  \bibinfo{journal}{Phys. Rev. D} \textbf{\bibinfo{volume}{66}},
  \bibinfo{pages}{084025} (\bibinfo{year}{2002}).

\bibitem[{\citenamefont{{Regimbau} and {de Freitas
  Pacheco}}(2006{\natexlab{b}})}]{2006A&A...447....1R}
\bibinfo{author}{\bibfnamefont{T.}~\bibnamefont{{Regimbau}}} \bibnamefont{and}
  \bibinfo{author}{\bibfnamefont{J.~A.} \bibnamefont{{de Freitas Pacheco}}},
  \bibinfo{journal}{Astronomy and Astrophysics} \textbf{\bibinfo{volume}{447}},
  \bibinfo{pages}{1} (\bibinfo{year}{2006}{\natexlab{b}}),
  \eprint{astro-ph/0509880}.

\bibitem[{\citenamefont{Regimbau and Mandic}(2008)}]{RegMan}
\bibinfo{author}{\bibfnamefont{T.}~\bibnamefont{Regimbau}} \bibnamefont{and}
  \bibinfo{author}{\bibfnamefont{V.}~\bibnamefont{Mandic}},
  \bibinfo{journal}{Class. Quant. Grav.} \textbf{\bibinfo{volume}{25}},
  \bibinfo{pages}{184018} (\bibinfo{year}{2008}).

\bibitem[{\citenamefont{{Marassi}
  et~al.}(2011{\natexlab{b}})\citenamefont{{Marassi}, {Ciolfi}, {Schneider},
  {Stella}, and {Ferrari}}}]{2011MNRAS.411.2549M}
\bibinfo{author}{\bibfnamefont{S.}~\bibnamefont{{Marassi}}},
  \bibinfo{author}{\bibfnamefont{R.}~\bibnamefont{{Ciolfi}}},
  \bibinfo{author}{\bibfnamefont{R.}~\bibnamefont{{Schneider}}},
  \bibinfo{author}{\bibfnamefont{L.}~\bibnamefont{{Stella}}}, \bibnamefont{and}
  \bibinfo{author}{\bibfnamefont{V.}~\bibnamefont{{Ferrari}}},
  \bibinfo{journal}{Monthly Notices of the Royal astronomical Society}
  \textbf{\bibinfo{volume}{411}}, \bibinfo{pages}{2549}
  (\bibinfo{year}{2011}{\natexlab{b}}), \eprint{1009.1240}.

\bibitem[{\citenamefont{{Howell} et~al.}(2011)\citenamefont{{Howell},
  {Regimbau}, {Corsi}, {Coward}, and {Burman}}}]{2011MNRAS.410.2123H}
\bibinfo{author}{\bibfnamefont{E.}~\bibnamefont{{Howell}}},
  \bibinfo{author}{\bibfnamefont{T.}~\bibnamefont{{Regimbau}}},
  \bibinfo{author}{\bibfnamefont{A.}~\bibnamefont{{Corsi}}},
  \bibinfo{author}{\bibfnamefont{D.}~\bibnamefont{{Coward}}}, \bibnamefont{and}
  \bibinfo{author}{\bibfnamefont{R.}~\bibnamefont{{Burman}}},
  \bibinfo{journal}{Monthly Notices of the Royal astronomical Society}
  \textbf{\bibinfo{volume}{410}}, \bibinfo{pages}{2123} (\bibinfo{year}{2011}),
  \eprint{1008.3941}.

\bibitem[{\citenamefont{{Wu} et~al.}(2013)\citenamefont{{Wu}, {Mandic}, and
  {Regimbau}}}]{2013PhRvD..87d2002W}
\bibinfo{author}{\bibfnamefont{C.-J.} \bibnamefont{{Wu}}},
  \bibinfo{author}{\bibfnamefont{V.}~\bibnamefont{{Mandic}}}, \bibnamefont{and}
  \bibinfo{author}{\bibfnamefont{T.}~\bibnamefont{{Regimbau}}},
  \bibinfo{journal}{\prd} \textbf{\bibinfo{volume}{87}}, \bibinfo{eid}{042002}
  (\bibinfo{year}{2013}).

\bibitem[{\citenamefont{Sandick et~al.}(2006)\citenamefont{Sandick, Olive,
  Daigne, and Vangioni}}]{firststars}
\bibinfo{author}{\bibfnamefont{P.}~\bibnamefont{Sandick}},
  \bibinfo{author}{\bibfnamefont{K.~A.} \bibnamefont{Olive}},
  \bibinfo{author}{\bibfnamefont{F.}~\bibnamefont{Daigne}}, \bibnamefont{and}
  \bibinfo{author}{\bibfnamefont{E.}~\bibnamefont{Vangioni}},
  \bibinfo{journal}{Phys. Rev. D} \textbf{\bibinfo{volume}{73}},
  \bibinfo{pages}{104024} (\bibinfo{year}{2006}).

\bibitem[{\citenamefont{Farmer and Phinney}(2003)}]{phinney_whitedwarfs}
\bibinfo{author}{\bibfnamefont{A.~J.} \bibnamefont{Farmer}} \bibnamefont{and}
  \bibinfo{author}{\bibfnamefont{E.}~\bibnamefont{Phinney}},
  \bibinfo{journal}{Mon. Not. Roy. Astron. Soc.}
  \textbf{\bibinfo{volume}{346}}, \bibinfo{pages}{1197} (\bibinfo{year}{2003}).

\bibitem[{\citenamefont{Abbott et~al.}(2005)}]{S3stoch}
\bibinfo{author}{\bibfnamefont{B.}~\bibnamefont{Abbott}} \bibnamefont{et~al.},
  \bibinfo{journal}{Phys. Rev. Lett.} \textbf{\bibinfo{volume}{95}},
  \bibinfo{pages}{221101} (\bibinfo{year}{2005}).

\bibitem[{\citenamefont{Abbott et~al.}(2007{\natexlab{a}})}]{S4stoch}
\bibinfo{author}{\bibfnamefont{B.}~\bibnamefont{Abbott}} \bibnamefont{et~al.},
  \bibinfo{journal}{Astrophys. J.} \textbf{\bibinfo{volume}{659}},
  \bibinfo{pages}{918} (\bibinfo{year}{2007}{\natexlab{a}}).

\bibitem[{\citenamefont{Abbott et~al.}(2009{\natexlab{a}})}]{S5stoch}
\bibinfo{author}{\bibfnamefont{B.}~\bibnamefont{Abbott}} \bibnamefont{et~al.},
  \bibinfo{journal}{Nature} \textbf{\bibinfo{volume}{460}},
  \bibinfo{pages}{990} (\bibinfo{year}{2009}{\natexlab{a}}).

\bibitem[{\citenamefont{Aasi et~al.}(2014)}]{S6stoch}
\bibinfo{author}{\bibfnamefont{J.}~\bibnamefont{Aasi}} \bibnamefont{et~al.},
  \bibinfo{journal}{Phys. Rev. Lett.} \textbf{\bibinfo{volume}{113}},
  \bibinfo{pages}{231101} (\bibinfo{year}{2014}).

\bibitem[{\citenamefont{Aasi et~al.}(2015)}]{H1H2stoch}
\bibinfo{author}{\bibfnamefont{J.}~\bibnamefont{Aasi}} \bibnamefont{et~al.},
  \bibinfo{journal}{Phys. Rev. D} \textbf{\bibinfo{volume}{91}},
  \bibinfo{pages}{022003} (\bibinfo{year}{2015}).

\bibitem[{\citenamefont{Abbott et~al.}(2007{\natexlab{b}})}]{S4radiometer}
\bibinfo{author}{\bibfnamefont{B.}~\bibnamefont{Abbott}} \bibnamefont{et~al.},
  \bibinfo{journal}{Phys. Rev. D} \textbf{\bibinfo{volume}{76}},
  \bibinfo{pages}{082003} (\bibinfo{year}{2007}{\natexlab{b}}).

\bibitem[{\citenamefont{Abbott et~al.}(2011)}]{S5SPH}
\bibinfo{author}{\bibfnamefont{B.}~\bibnamefont{Abbott}} \bibnamefont{et~al.},
  \bibinfo{journal}{Phys. Rev. Lett.} \textbf{\bibinfo{volume}{107}},
  \bibinfo{pages}{271102} (\bibinfo{year}{2011}).

\bibitem[{\citenamefont{Abbott et~al.}(2004)}]{LIGOS1}
\bibinfo{author}{\bibfnamefont{B.}~\bibnamefont{Abbott}} \bibnamefont{et~al.},
  \bibinfo{journal}{Nucl. Instr. Meth. A} \textbf{\bibinfo{volume}{517}},
  \bibinfo{pages}{154} (\bibinfo{year}{2004}).

\bibitem[{\citenamefont{Abbott et~al.}(2009{\natexlab{b}})}]{LIGOS5}
\bibinfo{author}{\bibfnamefont{B.}~\bibnamefont{Abbott}} \bibnamefont{et~al.},
  \bibinfo{journal}{Rep. Prog. Phys.} \textbf{\bibinfo{volume}{72}},
  \bibinfo{pages}{076901} (\bibinfo{year}{2009}{\natexlab{b}}).

\bibitem[{\citenamefont{Acernese et~al.}(2006)}]{Virgo1}
\bibinfo{author}{\bibfnamefont{F.}~\bibnamefont{Acernese}}
  \bibnamefont{et~al.}, \bibinfo{journal}{Class. Quant. Grav.}
  \textbf{\bibinfo{volume}{23}}, \bibinfo{pages}{S63} (\bibinfo{year}{2006}).

\bibitem[{\citenamefont{Mandic et~al.}(2012)\citenamefont{Mandic, Thrane,
  Giampanis, and Regimbau}}]{paramest}
\bibinfo{author}{\bibfnamefont{V.}~\bibnamefont{Mandic}},
  \bibinfo{author}{\bibfnamefont{E.}~\bibnamefont{Thrane}},
  \bibinfo{author}{\bibfnamefont{S.}~\bibnamefont{Giampanis}},
  \bibnamefont{and} \bibinfo{author}{\bibfnamefont{T.}~\bibnamefont{Regimbau}},
  \bibinfo{journal}{Phys. Rev. Lett.} \textbf{\bibinfo{volume}{109}},
  \bibinfo{pages}{171102} (\bibinfo{year}{2012}).

\bibitem[{\citenamefont{Crowder et~al.}(2013)\citenamefont{Crowder, Namba,
  Mandic, Mukhoyama, and Peloso}}]{parviol}
\bibinfo{author}{\bibfnamefont{S.~G.} \bibnamefont{Crowder}},
  \bibinfo{author}{\bibfnamefont{R.}~\bibnamefont{Namba}},
  \bibinfo{author}{\bibfnamefont{V.}~\bibnamefont{Mandic}},
  \bibinfo{author}{\bibfnamefont{S.}~\bibnamefont{Mukhoyama}},
  \bibnamefont{and} \bibinfo{author}{\bibfnamefont{M.}~\bibnamefont{Peloso}},
  \bibinfo{journal}{Phys. Lett. B} \textbf{\bibinfo{volume}{726}},
  \bibinfo{pages}{66} (\bibinfo{year}{2013}).

\bibitem[{\citenamefont{{G.~M. Harry (for the LIGO Scientific
  Collaboration)}}(2010)}]{aLIGO2}
\bibinfo{author}{\bibnamefont{{G.~M. Harry (for the LIGO Scientific
  Collaboration)}}}, \bibinfo{journal}{Class. Quant. Grav.}
  \textbf{\bibinfo{volume}{27}}, \bibinfo{pages}{084006}
  (\bibinfo{year}{2010}).

\bibitem[{aVi()}]{aVirgo}
\bibinfo{howpublished}{\url{https://wwwcascina.virgo.infn.it/advirgo/docs.html}}.

\bibitem[{\citenamefont{Willke et~al.}(2006)}]{GEOHF}
\bibinfo{author}{\bibfnamefont{B.}~\bibnamefont{Willke}} \bibnamefont{et~al.},
  \bibinfo{journal}{Class. Quant. Grav.} \textbf{\bibinfo{volume}{23}},
  \bibinfo{pages}{S207} (\bibinfo{year}{2006}).

\bibitem[{\citenamefont{Affeldt et~al.}(2014)}]{GEOHF2}
\bibinfo{author}{\bibfnamefont{C.}~\bibnamefont{Affeldt}} \bibnamefont{et~al.},
  \bibinfo{journal}{Class. Quant. Grav.} \textbf{\bibinfo{volume}{31}},
  \bibinfo{pages}{224002} (\bibinfo{year}{2014}).

\bibitem[{\citenamefont{{Y. Aso and Y. Michimura and K. Somiya and M. Ando and
  O. Miyakawa and T. Sekiguchi and D. Tatsumi and and H.
  Yamamoto}}(2013)}]{KAGRA1}
\bibinfo{author}{\bibnamefont{{Y. Aso and Y. Michimura and K. Somiya and M.
  Ando and O. Miyakawa and T. Sekiguchi and D. Tatsumi and and H. Yamamoto}}},
  \bibinfo{journal}{Phys. Rev. D} \textbf{\bibinfo{volume}{88}},
  \bibinfo{pages}{043007} (\bibinfo{year}{2013}).

\bibitem[{\citenamefont{(for~the KAGRA~collaboration)}(2012)}]{KAGRA2}
\bibinfo{author}{\bibfnamefont{K.~S.} \bibnamefont{(for~the
  KAGRA~collaboration)}}, \bibinfo{journal}{Class. Quant. Grav.}
  \textbf{\bibinfo{volume}{29}}, \bibinfo{pages}{124007}
  (\bibinfo{year}{2012}).

\bibitem[{\citenamefont{science team}(2011)}]{ET}
\bibinfo{author}{\bibfnamefont{T.~E.} \bibnamefont{science team}},
  \bibinfo{journal}{https://tds.ego-gw.it/ql/?c=7954}  (\bibinfo{year}{2011}).

\bibitem[{\citenamefont{Ott et~al.}(2011)\citenamefont{Ott, Reisswig,
  Schnetter, O'Connor, Sperhake, L\"offler, Diener, Abdikamalov, Hawke, and
  Burrows}}]{OttPRL}
\bibinfo{author}{\bibfnamefont{C.}~\bibnamefont{Ott}},
  \bibinfo{author}{\bibfnamefont{C.}~\bibnamefont{Reisswig}},
  \bibinfo{author}{\bibfnamefont{E.}~\bibnamefont{Schnetter}},
  \bibinfo{author}{\bibfnamefont{E.}~\bibnamefont{O'Connor}},
  \bibinfo{author}{\bibfnamefont{U.}~\bibnamefont{Sperhake}},
  \bibinfo{author}{\bibfnamefont{F.}~\bibnamefont{L\"offler}},
  \bibinfo{author}{\bibfnamefont{P.}~\bibnamefont{Diener}},
  \bibinfo{author}{\bibfnamefont{E.}~\bibnamefont{Abdikamalov}},
  \bibinfo{author}{\bibfnamefont{I.}~\bibnamefont{Hawke}}, \bibnamefont{and}
  \bibinfo{author}{\bibfnamefont{A.}~\bibnamefont{Burrows}},
  \bibinfo{journal}{Phys. Rev. Lett.} \textbf{\bibinfo{volume}{106}},
  \bibinfo{pages}{161103} (\bibinfo{year}{2011}).

\bibitem[{\citenamefont{Ott et~al.}(2013)}]{ott2013}
\bibinfo{author}{\bibfnamefont{C.}~\bibnamefont{Ott}} \bibnamefont{et~al.},
  \bibinfo{journal}{Astrophys. J.} \textbf{\bibinfo{volume}{768}},
  \bibinfo{pages}{115} (\bibinfo{year}{2013}).

\bibitem[{\citenamefont{Muller et~al.}(2013)}]{muller2013}
\bibinfo{author}{\bibfnamefont{B.}~\bibnamefont{Muller}} \bibnamefont{et~al.},
  \bibinfo{journal}{Astrophys. J.} \textbf{\bibinfo{volume}{768}},
  \bibinfo{pages}{115} (\bibinfo{year}{2013}).

\bibitem[{\citenamefont{{Ferrari}
  et~al.}(1999{\natexlab{b}})\citenamefont{{Ferrari}, {Matarrese}, and
  {Schneider}}}]{ferrari_ccbh}
\bibinfo{author}{\bibfnamefont{V.}~\bibnamefont{{Ferrari}}},
  \bibinfo{author}{\bibfnamefont{S.}~\bibnamefont{{Matarrese}}},
  \bibnamefont{and}
  \bibinfo{author}{\bibfnamefont{R.}~\bibnamefont{{Schneider}}},
  \bibinfo{journal}{Mon. Not. Roy. Astron. Soc.}
  \textbf{\bibinfo{volume}{303}}, \bibinfo{pages}{247}
  (\bibinfo{year}{1999}{\natexlab{b}}), \eprint{astro-ph/9804259}.

\bibitem[{\citenamefont{{de Araujo} et~al.}(2000)\citenamefont{{de Araujo},
  {Miranda}, and {Aguiar}}}]{2000PhRvD..61l4015D}
\bibinfo{author}{\bibfnamefont{J.~C.~N.} \bibnamefont{{de Araujo}}},
  \bibinfo{author}{\bibfnamefont{O.~D.} \bibnamefont{{Miranda}}},
  \bibnamefont{and} \bibinfo{author}{\bibfnamefont{O.~D.}
  \bibnamefont{{Aguiar}}}, \bibinfo{journal}{Phys. Rev. D}
  \textbf{\bibinfo{volume}{61}}, \bibinfo{eid}{124015} (\bibinfo{year}{2000}),
  \eprint{astro-ph/0004395}.

\bibitem[{\citenamefont{{de Araujo} et~al.}(2002)\citenamefont{{de Araujo},
  {Miranda}, and {Aguiar}}}]{2002MNRAS.330..651D}
\bibinfo{author}{\bibfnamefont{J.~C.~N.} \bibnamefont{{de Araujo}}},
  \bibinfo{author}{\bibfnamefont{O.~D.} \bibnamefont{{Miranda}}},
  \bibnamefont{and} \bibinfo{author}{\bibfnamefont{O.~D.}
  \bibnamefont{{Aguiar}}}, \bibinfo{journal}{Mon. Not. Roy. Astron. Soc.}
  \textbf{\bibinfo{volume}{330}}, \bibinfo{pages}{651} (\bibinfo{year}{2002}),
  \eprint{astro-ph/0202037}.

\bibitem[{\citenamefont{de~Araujo et~al.}(2002)\citenamefont{de~Araujo,
  Miranda, and Aguiar}}]{araujo}
\bibinfo{author}{\bibfnamefont{J.}~\bibnamefont{de~Araujo}},
  \bibinfo{author}{\bibfnamefont{O.}~\bibnamefont{Miranda}}, \bibnamefont{and}
  \bibinfo{author}{\bibfnamefont{O.}~\bibnamefont{Aguiar}},
  \bibinfo{journal}{Class. Quant. Grav.} \textbf{\bibinfo{volume}{19}},
  \bibinfo{pages}{1335} (\bibinfo{year}{2002}).

\bibitem[{\citenamefont{{de Araujo}
  et~al.}(2004{\natexlab{a}})\citenamefont{{de Araujo}, {Miranda}, and
  {Aguiar}}}]{2004MNRAS.348.1373D}
\bibinfo{author}{\bibfnamefont{J.~C.~N.} \bibnamefont{{de Araujo}}},
  \bibinfo{author}{\bibfnamefont{O.~D.} \bibnamefont{{Miranda}}},
  \bibnamefont{and} \bibinfo{author}{\bibfnamefont{O.~D.}
  \bibnamefont{{Aguiar}}}, \bibinfo{journal}{mnras}
  \textbf{\bibinfo{volume}{348}}, \bibinfo{pages}{1373}
  (\bibinfo{year}{2004}{\natexlab{a}}).

\bibitem[{\citenamefont{{de Araujo}
  et~al.}(2004{\natexlab{b}})\citenamefont{{de Araujo}, {Miranda}, and
  {Aguiar}}}]{2004CQGra..21S.545D}
\bibinfo{author}{\bibfnamefont{J.~C.~N.} \bibnamefont{{de Araujo}}},
  \bibinfo{author}{\bibfnamefont{O.~D.} \bibnamefont{{Miranda}}},
  \bibnamefont{and} \bibinfo{author}{\bibfnamefont{O.~D.}
  \bibnamefont{{Aguiar}}}, \bibinfo{journal}{Classical and Quantum Gravity}
  \textbf{\bibinfo{volume}{21}}, \bibinfo{pages}{545}
  (\bibinfo{year}{2004}{\natexlab{b}}).

\bibitem[{\citenamefont{Buonanno et~al.}(2005)\citenamefont{Buonanno, Sigl,
  Raffelt, Janka, and Muller}}]{buonanno}
\bibinfo{author}{\bibfnamefont{A.}~\bibnamefont{Buonanno}},
  \bibinfo{author}{\bibfnamefont{G.}~\bibnamefont{Sigl}},
  \bibinfo{author}{\bibfnamefont{G.}~\bibnamefont{Raffelt}},
  \bibinfo{author}{\bibfnamefont{H.}~\bibnamefont{Janka}}, \bibnamefont{and}
  \bibinfo{author}{\bibfnamefont{E.}~\bibnamefont{Muller}},
  \bibinfo{journal}{Phys. Rev. D} \textbf{\bibinfo{volume}{72}},
  \bibinfo{pages}{084001} (\bibinfo{year}{2005}).

\bibitem[{\citenamefont{Coward et~al.}(2002)\citenamefont{Coward, Burman, and
  Blair}}]{coward}
\bibinfo{author}{\bibfnamefont{D.}~\bibnamefont{Coward}},
  \bibinfo{author}{\bibfnamefont{R.}~\bibnamefont{Burman}}, \bibnamefont{and}
  \bibinfo{author}{\bibfnamefont{D.}~\bibnamefont{Blair}},
  \bibinfo{journal}{Mon. Not. Roy. Astron. Soc.}
  \textbf{\bibinfo{volume}{329}}, \bibinfo{pages}{411} (\bibinfo{year}{2002}).

\bibitem[{\citenamefont{{Zhu} et~al.}(2010)\citenamefont{{Zhu}, {Howell}, and
  {Blair}}}]{2010MNRAS.409L.132Z}
\bibinfo{author}{\bibfnamefont{X.-J.} \bibnamefont{{Zhu}}},
  \bibinfo{author}{\bibfnamefont{E.}~\bibnamefont{{Howell}}}, \bibnamefont{and}
  \bibinfo{author}{\bibfnamefont{D.}~\bibnamefont{{Blair}}},
  \bibinfo{journal}{Monthly Notices of the Royal Astronomical Society}
  \textbf{\bibinfo{volume}{409}}, \bibinfo{pages}{L132} (\bibinfo{year}{2010}),
  \eprint{1008.0472}.

\bibitem[{\citenamefont{Marassi et~al.}(2009)\citenamefont{Marassi, Schneider,
  and Ferrari}}]{marassi_cc}
\bibinfo{author}{\bibfnamefont{S.}~\bibnamefont{Marassi}},
  \bibinfo{author}{\bibfnamefont{R.}~\bibnamefont{Schneider}},
  \bibnamefont{and} \bibinfo{author}{\bibfnamefont{V.}~\bibnamefont{Ferrari}},
  \bibinfo{journal}{Mon. Not. Roy. Astron. Soc.}
  \textbf{\bibinfo{volume}{398}}, \bibinfo{pages}{293} (\bibinfo{year}{2009}).

\bibitem[{\citenamefont{Pacucci et~al.}(2015)\citenamefont{Pacucci, Ferrara,
  and Marassi}}]{pacucci}
\bibinfo{author}{\bibfnamefont{F.}~\bibnamefont{Pacucci}},
  \bibinfo{author}{\bibfnamefont{A.}~\bibnamefont{Ferrara}}, \bibnamefont{and}
  \bibinfo{author}{\bibfnamefont{S.}~\bibnamefont{Marassi}},
  \bibinfo{journal}{arXiv:1502.04125}  (\bibinfo{year}{2015}).

\bibitem[{\citenamefont{Echeverria}(1989)}]{echeveria}
\bibinfo{author}{\bibfnamefont{F.}~\bibnamefont{Echeverria}},
  \bibinfo{journal}{Phys. Rev. D} \textbf{\bibinfo{volume}{40}},
  \bibinfo{pages}{3194} (\bibinfo{year}{1989}).

\bibitem[{\citenamefont{Stark and Piran}(1985)}]{SP1}
\bibinfo{author}{\bibfnamefont{R.}~\bibnamefont{Stark}} \bibnamefont{and}
  \bibinfo{author}{\bibfnamefont{T.}~\bibnamefont{Piran}},
  \bibinfo{journal}{Phys. Rev. Lett.} \textbf{\bibinfo{volume}{55}},
  \bibinfo{pages}{891} (\bibinfo{year}{1985}).

\bibitem[{\citenamefont{Belczynski et~al.}()}]{startrack}
\bibinfo{author}{\bibfnamefont{K.}~\bibnamefont{Belczynski}}
  \bibnamefont{et~al.},
  \bibinfo{howpublished}{\url{http://www.syntheticuniverse.org}}.

\bibitem[{\citenamefont{Lilly et~al.}(1996)\citenamefont{Lilly, Fevre, Hammer,
  and Crampton}}]{lilly}
\bibinfo{author}{\bibfnamefont{S.}~\bibnamefont{Lilly}},
  \bibinfo{author}{\bibfnamefont{O.~L.} \bibnamefont{Fevre}},
  \bibinfo{author}{\bibfnamefont{F.}~\bibnamefont{Hammer}}, \bibnamefont{and}
  \bibinfo{author}{\bibfnamefont{D.}~\bibnamefont{Crampton}},
  \bibinfo{journal}{Astrop. J.} \textbf{\bibinfo{volume}{460}},
  \bibinfo{pages}{L1} (\bibinfo{year}{1996}).

\bibitem[{\citenamefont{Hopkins and Beacom}(2006)}]{hopkins}
\bibinfo{author}{\bibfnamefont{A.~M.} \bibnamefont{Hopkins}} \bibnamefont{and}
  \bibinfo{author}{\bibfnamefont{J.}~\bibnamefont{Beacom}},
  \bibinfo{journal}{Astrop. J.} \textbf{\bibinfo{volume}{651}},
  \bibinfo{pages}{142} (\bibinfo{year}{2006}).

\bibitem[{\citenamefont{Fardal et~al.}(2007)\citenamefont{Fardal, Katz,
  Weinberg, and Davé}}]{fardal}
\bibinfo{author}{\bibfnamefont{M.~A.} \bibnamefont{Fardal}},
  \bibinfo{author}{\bibfnamefont{N.}~\bibnamefont{Katz}},
  \bibinfo{author}{\bibfnamefont{D.~H.} \bibnamefont{Weinberg}},
  \bibnamefont{and} \bibinfo{author}{\bibfnamefont{R.}~\bibnamefont{Davé}},
  \bibinfo{journal}{Mon. Not. Roy. Astron. Soc.}
  \textbf{\bibinfo{volume}{379}}, \bibinfo{pages}{985} (\bibinfo{year}{2007}).

\bibitem[{\citenamefont{Wilkins et~al.}(2008)\citenamefont{Wilkins, Trentham,
  and Hopkins}}]{wilkins}
\bibinfo{author}{\bibfnamefont{S.~M.} \bibnamefont{Wilkins}},
  \bibinfo{author}{\bibfnamefont{N.}~\bibnamefont{Trentham}}, \bibnamefont{and}
  \bibinfo{author}{\bibfnamefont{A.}~\bibnamefont{Hopkins}},
  \bibinfo{journal}{\rm arXiv:0803.4024}  (\bibinfo{year}{2008}).

\bibitem[{\citenamefont{Nagamine et~al.}(2006)\citenamefont{Nagamine, Ostriker,
  Fukugita, and Cen}}]{nagamine}
\bibinfo{author}{\bibfnamefont{K.}~\bibnamefont{Nagamine}},
  \bibinfo{author}{\bibfnamefont{J.~P.} \bibnamefont{Ostriker}},
  \bibinfo{author}{\bibfnamefont{M.}~\bibnamefont{Fukugita}}, \bibnamefont{and}
  \bibinfo{author}{\bibfnamefont{R.}~\bibnamefont{Cen}},
  \bibinfo{journal}{Astrop. J.} \textbf{\bibinfo{volume}{653}},
  \bibinfo{pages}{881} (\bibinfo{year}{2006}).

\bibitem[{\citenamefont{Hernquist and Springel}(2003)}]{springel}
\bibinfo{author}{\bibfnamefont{L.}~\bibnamefont{Hernquist}} \bibnamefont{and}
  \bibinfo{author}{\bibfnamefont{V.}~\bibnamefont{Springel}},
  \bibinfo{journal}{Mon. Not. Roy. Astron. Soc.}
  \textbf{\bibinfo{volume}{341}}, \bibinfo{pages}{1253} (\bibinfo{year}{2003}).

\bibitem[{\citenamefont{Behroozi et~al.}(2013)\citenamefont{Behroozi, Wechsler,
  and Conroy}}]{behroozi}
\bibinfo{author}{\bibfnamefont{P.}~\bibnamefont{Behroozi}},
  \bibinfo{author}{\bibfnamefont{R.}~\bibnamefont{Wechsler}}, \bibnamefont{and}
  \bibinfo{author}{\bibfnamefont{C.}~\bibnamefont{Conroy}},
  \bibinfo{journal}{Astrophys. J.} \textbf{\bibinfo{volume}{770}},
  \bibinfo{pages}{57} (\bibinfo{year}{2013}).

\bibitem[{\citenamefont{Oesch et~al.}(2014{\natexlab{a}})}]{oesch1}
\bibinfo{author}{\bibfnamefont{P.}~\bibnamefont{Oesch}} \bibnamefont{et~al.},
  \bibinfo{journal}{Astrophys. J.} \textbf{\bibinfo{volume}{786}},
  \bibinfo{pages}{108} (\bibinfo{year}{2014}{\natexlab{a}}).

\bibitem[{\citenamefont{Oesch et~al.}(2014{\natexlab{b}})}]{oesch2}
\bibinfo{author}{\bibfnamefont{P.}~\bibnamefont{Oesch}} \bibnamefont{et~al.},
  \bibinfo{journal}{arXiv:1409.1228}  (\bibinfo{year}{2014}{\natexlab{b}}).

\bibitem[{\citenamefont{Robertson and Ellis}(2012)}]{robertsonellis}
\bibinfo{author}{\bibfnamefont{B.}~\bibnamefont{Robertson}} \bibnamefont{and}
  \bibinfo{author}{\bibfnamefont{R.}~\bibnamefont{Ellis}},
  \bibinfo{journal}{Astrophys. J.} \textbf{\bibinfo{volume}{744}},
  \bibinfo{pages}{95} (\bibinfo{year}{2012}).

\bibitem[{\citenamefont{Wang}(2013)}]{wang}
\bibinfo{author}{\bibfnamefont{F.~Y.} \bibnamefont{Wang}},
  \bibinfo{journal}{A\&A} \textbf{\bibinfo{volume}{556}}, \bibinfo{pages}{A90}
  (\bibinfo{year}{2013}).

\bibitem[{\citenamefont{Kistler et~al.}(2013)\citenamefont{Kistler, Yuksel, and
  Hopkins}}]{kistler}
\bibinfo{author}{\bibfnamefont{M.}~\bibnamefont{Kistler}},
  \bibinfo{author}{\bibfnamefont{H.}~\bibnamefont{Yuksel}}, \bibnamefont{and}
  \bibinfo{author}{\bibfnamefont{A.}~\bibnamefont{Hopkins}},
  \bibinfo{journal}{arXiv:1305.1630}  (\bibinfo{year}{2013}).

\bibitem[{\citenamefont{Trenti et~al.}(2013)\citenamefont{Trenti, Perna, and
  Tacchella}}]{trenti}
\bibinfo{author}{\bibfnamefont{M.}~\bibnamefont{Trenti}},
  \bibinfo{author}{\bibfnamefont{R.}~\bibnamefont{Perna}}, \bibnamefont{and}
  \bibinfo{author}{\bibfnamefont{S.}~\bibnamefont{Tacchella}},
  \bibinfo{journal}{Astrophys. J. Lett.} \textbf{\bibinfo{volume}{773}},
  \bibinfo{pages}{22} (\bibinfo{year}{2013}).

\bibitem[{\citenamefont{Behroozi and Silk}(2015)}]{behroozisilk}
\bibinfo{author}{\bibfnamefont{P.}~\bibnamefont{Behroozi}} \bibnamefont{and}
  \bibinfo{author}{\bibfnamefont{J.}~\bibnamefont{Silk}},
  \bibinfo{journal}{Astrophys. J.} \textbf{\bibinfo{volume}{799}},
  \bibinfo{pages}{32} (\bibinfo{year}{2015}).

\bibitem[{\citenamefont{Vangioni et~al.}(2014)\citenamefont{Vangioni, Olive,
  Prestegard, Silk, Petitjean, and Mandic}}]{vangioni}
\bibinfo{author}{\bibfnamefont{E.}~\bibnamefont{Vangioni}},
  \bibinfo{author}{\bibfnamefont{K.}~\bibnamefont{Olive}},
  \bibinfo{author}{\bibfnamefont{T.}~\bibnamefont{Prestegard}},
  \bibinfo{author}{\bibfnamefont{J.}~\bibnamefont{Silk}},
  \bibinfo{author}{\bibfnamefont{P.}~\bibnamefont{Petitjean}},
  \bibnamefont{and} \bibinfo{author}{\bibfnamefont{V.}~\bibnamefont{Mandic}},
  \bibinfo{journal}{arXiv:1409.2462}  (\bibinfo{year}{2014}).

\bibitem[{\citenamefont{Hinshaw et~al.}(2013)}]{WMAP}
\bibinfo{author}{\bibfnamefont{G.}~\bibnamefont{Hinshaw}} \bibnamefont{et~al.},
  \bibinfo{journal}{Astrophys. J. Supl} \textbf{\bibinfo{volume}{208}},
  \bibinfo{pages}{19} (\bibinfo{year}{2013}).

\bibitem[{\citenamefont{{Crowther} et~al.}(2010)\citenamefont{{Crowther},
  {Schnurr}, {Hirschi}, {Yusof}, {Parker}, {Goodwin}, and
  {Kassim}}}]{Crowther2010}
\bibinfo{author}{\bibfnamefont{P.~A.} \bibnamefont{{Crowther}}},
  \bibinfo{author}{\bibfnamefont{O.}~\bibnamefont{{Schnurr}}},
  \bibinfo{author}{\bibfnamefont{R.}~\bibnamefont{{Hirschi}}},
  \bibinfo{author}{\bibfnamefont{N.}~\bibnamefont{{Yusof}}},
  \bibinfo{author}{\bibfnamefont{R.~J.} \bibnamefont{{Parker}}},
  \bibinfo{author}{\bibfnamefont{S.~P.} \bibnamefont{{Goodwin}}},
  \bibnamefont{and} \bibinfo{author}{\bibfnamefont{H.~A.}
  \bibnamefont{{Kassim}}}, \bibinfo{journal}{Mon. Not. Roy. Astron. Soc.}
  \textbf{\bibinfo{volume}{408}}, \bibinfo{pages}{731} (\bibinfo{year}{2010}),
  \eprint{1007.3284}.

\bibitem[{aLI()}]{aLIGOsens}
\bibinfo{howpublished}{\url{https://dcc.ligo.org/LIGO-T0900288/public}}.

\bibitem[{\citenamefont{Belczynski et~al.}(2010)\citenamefont{Belczynski,
  Bulik, Fryer, Ruiter, Valsecchi, Vink, and Hurley}}]{belczynski}
\bibinfo{author}{\bibfnamefont{K.}~\bibnamefont{Belczynski}},
  \bibinfo{author}{\bibfnamefont{T.}~\bibnamefont{Bulik}},
  \bibinfo{author}{\bibfnamefont{C.}~\bibnamefont{Fryer}},
  \bibinfo{author}{\bibfnamefont{A.}~\bibnamefont{Ruiter}},
  \bibinfo{author}{\bibfnamefont{F.}~\bibnamefont{Valsecchi}},
  \bibinfo{author}{\bibfnamefont{J.}~\bibnamefont{Vink}}, \bibnamefont{and}
  \bibinfo{author}{\bibfnamefont{J.}~\bibnamefont{Hurley}},
  \bibinfo{journal}{Astrophys. J.} \textbf{\bibinfo{volume}{714}},
  \bibinfo{pages}{1217} (\bibinfo{year}{2010}).

\bibitem[{\citenamefont{Kobayashi and Meszaros}(2003)}]{kobayashi}
\bibinfo{author}{\bibfnamefont{S.}~\bibnamefont{Kobayashi}} \bibnamefont{and}
  \bibinfo{author}{\bibfnamefont{P.}~\bibnamefont{Meszaros}},
  \bibinfo{journal}{Astrophys. J.} \textbf{\bibinfo{volume}{589}},
  \bibinfo{pages}{861} (\bibinfo{year}{2003}).

\bibitem[{\citenamefont{Baiotti and Rezzolla}(2006)}]{baiotti}
\bibinfo{author}{\bibfnamefont{L.}~\bibnamefont{Baiotti}} \bibnamefont{and}
  \bibinfo{author}{\bibfnamefont{L.}~\bibnamefont{Rezzolla}},
  \bibinfo{journal}{Phys. Rev. Lett.} \textbf{\bibinfo{volume}{97}},
  \bibinfo{pages}{141101} (\bibinfo{year}{2006}).

\bibitem[{\citenamefont{Woosley and Weaver}(1995)}]{ww95}
\bibinfo{author}{\bibfnamefont{S.}~\bibnamefont{Woosley}} \bibnamefont{and}
  \bibinfo{author}{\bibfnamefont{T.}~\bibnamefont{Weaver}},
  \bibinfo{journal}{Astrophys. J. Supl} \textbf{\bibinfo{volume}{101}},
  \bibinfo{pages}{181} (\bibinfo{year}{1995}).

\bibitem[{\citenamefont{{Belczynski} et~al.}(2002)\citenamefont{{Belczynski},
  {Kalogera}, and {Bulik}}}]{Belczynski2002}
\bibinfo{author}{\bibfnamefont{K.}~\bibnamefont{{Belczynski}}},
  \bibinfo{author}{\bibfnamefont{V.}~\bibnamefont{{Kalogera}}},
  \bibnamefont{and} \bibinfo{author}{\bibfnamefont{T.}~\bibnamefont{{Bulik}}},
  \bibinfo{journal}{Astrophys. J.} \textbf{\bibinfo{volume}{572}},
  \bibinfo{pages}{407} (\bibinfo{year}{2002}), \eprint{arXiv:astro-ph/0111452}.

\bibitem[{\citenamefont{{Belczynski} et~al.}(2008)\citenamefont{{Belczynski},
  {Kalogera}, {Rasio}, {Taam}, {Zezas}, {Bulik}, {Maccarone}, and
  {Ivanova}}}]{Belczynski2008}
\bibinfo{author}{\bibfnamefont{K.}~\bibnamefont{{Belczynski}}},
  \bibinfo{author}{\bibfnamefont{V.}~\bibnamefont{{Kalogera}}},
  \bibinfo{author}{\bibfnamefont{F.~A.} \bibnamefont{{Rasio}}},
  \bibinfo{author}{\bibfnamefont{R.~E.} \bibnamefont{{Taam}}},
  \bibinfo{author}{\bibfnamefont{A.}~\bibnamefont{{Zezas}}},
  \bibinfo{author}{\bibfnamefont{T.}~\bibnamefont{{Bulik}}},
  \bibinfo{author}{\bibfnamefont{T.~J.} \bibnamefont{{Maccarone}}},
  \bibnamefont{and}
  \bibinfo{author}{\bibfnamefont{N.}~\bibnamefont{{Ivanova}}},
  \bibinfo{journal}{Astrophys. J. Supl} \textbf{\bibinfo{volume}{174}},
  \bibinfo{pages}{223} (\bibinfo{year}{2008}), \eprint{arXiv:astro-ph/0511811}.

\bibitem[{\citenamefont{{Hurley} et~al.}(2000)\citenamefont{{Hurley}, {Pols},
  and {Tout}}}]{Hurley2000}
\bibinfo{author}{\bibfnamefont{J.~R.} \bibnamefont{{Hurley}}},
  \bibinfo{author}{\bibfnamefont{O.~R.} \bibnamefont{{Pols}}},
  \bibnamefont{and} \bibinfo{author}{\bibfnamefont{C.~A.}
  \bibnamefont{{Tout}}}, \bibinfo{journal}{Mon. Not. Roy. Astron. Soc.}
  \textbf{\bibinfo{volume}{315}}, \bibinfo{pages}{543} (\bibinfo{year}{2000}),
  \eprint{arXiv:astro-ph/0001295}.

\bibitem[{\citenamefont{{Dominik} et~al.}(2012)\citenamefont{{Dominik},
  {Belczynski}, {Fryer}, {Holz}, {Berti}, {Bulik}, {Mandel}, and
  {O'Shaughnessy}}}]{Dominik2012}
\bibinfo{author}{\bibfnamefont{M.}~\bibnamefont{{Dominik}}},
  \bibinfo{author}{\bibfnamefont{K.}~\bibnamefont{{Belczynski}}},
  \bibinfo{author}{\bibfnamefont{C.}~\bibnamefont{{Fryer}}},
  \bibinfo{author}{\bibfnamefont{D.~E.} \bibnamefont{{Holz}}},
  \bibinfo{author}{\bibfnamefont{E.}~\bibnamefont{{Berti}}},
  \bibinfo{author}{\bibfnamefont{T.}~\bibnamefont{{Bulik}}},
  \bibinfo{author}{\bibfnamefont{I.}~\bibnamefont{{Mandel}}}, \bibnamefont{and}
  \bibinfo{author}{\bibfnamefont{R.}~\bibnamefont{{O'Shaughnessy}}},
  \bibinfo{journal}{Astrophys. J.} \textbf{\bibinfo{volume}{759}},
  \bibinfo{eid}{52} (\bibinfo{year}{2012}), \eprint{1202.4901}.

\bibitem[{\citenamefont{{Sana} et~al.}(2012)\citenamefont{{Sana}, {de Mink},
  {de Koter}, {Langer}, {Evans}, {Gieles}, {Gosset}, {Izzard}, {Le Bouquin},
  and {Schneider}}}]{Sana2012}
\bibinfo{author}{\bibfnamefont{H.}~\bibnamefont{{Sana}}},
  \bibinfo{author}{\bibfnamefont{S.~E.} \bibnamefont{{de Mink}}},
  \bibinfo{author}{\bibfnamefont{A.}~\bibnamefont{{de Koter}}},
  \bibinfo{author}{\bibfnamefont{N.}~\bibnamefont{{Langer}}},
  \bibinfo{author}{\bibfnamefont{C.~J.} \bibnamefont{{Evans}}},
  \bibinfo{author}{\bibfnamefont{M.}~\bibnamefont{{Gieles}}},
  \bibinfo{author}{\bibfnamefont{E.}~\bibnamefont{{Gosset}}},
  \bibinfo{author}{\bibfnamefont{R.~G.} \bibnamefont{{Izzard}}},
  \bibinfo{author}{\bibfnamefont{J.-B.} \bibnamefont{{Le Bouquin}}},
  \bibnamefont{and} \bibinfo{author}{\bibfnamefont{F.~R.~N.}
  \bibnamefont{{Schneider}}}, \bibinfo{journal}{Science}
  \textbf{\bibinfo{volume}{337}}, \bibinfo{pages}{444} (\bibinfo{year}{2012}),
  \eprint{1207.6397}.

\bibitem[{\citenamefont{{Vink} et~al.}(2001)\citenamefont{{Vink}, {de Koter},
  and {Lamers}}}]{Vink2001}
\bibinfo{author}{\bibfnamefont{J.~S.} \bibnamefont{{Vink}}},
  \bibinfo{author}{\bibfnamefont{A.}~\bibnamefont{{de Koter}}},
  \bibnamefont{and} \bibinfo{author}{\bibfnamefont{H.~J.~G.~L.~M.}
  \bibnamefont{{Lamers}}}, \bibinfo{journal}{A\&A}
  \textbf{\bibinfo{volume}{369}}, \bibinfo{pages}{574} (\bibinfo{year}{2001}),
  \eprint{astro-ph/0101509}.

\bibitem[{\citenamefont{{Fryer} et~al.}(2012)\citenamefont{{Fryer},
  {Belczynski}, {Wiktorowicz}, {Dominik}, {Kalogera}, and {Holz}}}]{Fryer2012}
\bibinfo{author}{\bibfnamefont{C.~L.} \bibnamefont{{Fryer}}},
  \bibinfo{author}{\bibfnamefont{K.}~\bibnamefont{{Belczynski}}},
  \bibinfo{author}{\bibfnamefont{G.}~\bibnamefont{{Wiktorowicz}}},
  \bibinfo{author}{\bibfnamefont{M.}~\bibnamefont{{Dominik}}},
  \bibinfo{author}{\bibfnamefont{V.}~\bibnamefont{{Kalogera}}},
  \bibnamefont{and} \bibinfo{author}{\bibfnamefont{D.~E.}
  \bibnamefont{{Holz}}}, \bibinfo{journal}{Astrophys. J.}
  \textbf{\bibinfo{volume}{749}}, \bibinfo{eid}{91} (\bibinfo{year}{2012}),
  \eprint{1110.1726}.

\bibitem[{\citenamefont{{Belczynski} et~al.}(2012)\citenamefont{{Belczynski},
  {Wiktorowicz}, {Fryer}, {Holz}, and {Kalogera}}}]{Belczynski2012}
\bibinfo{author}{\bibfnamefont{K.}~\bibnamefont{{Belczynski}}},
  \bibinfo{author}{\bibfnamefont{G.}~\bibnamefont{{Wiktorowicz}}},
  \bibinfo{author}{\bibfnamefont{C.~L.} \bibnamefont{{Fryer}}},
  \bibinfo{author}{\bibfnamefont{D.~E.} \bibnamefont{{Holz}}},
  \bibnamefont{and}
  \bibinfo{author}{\bibfnamefont{V.}~\bibnamefont{{Kalogera}}},
  \bibinfo{journal}{Astrophys. J.} \textbf{\bibinfo{volume}{757}},
  \bibinfo{eid}{91} (\bibinfo{year}{2012}), \eprint{1110.1635}.

\bibitem[{\citenamefont{{{\"O}zel} et~al.}(2010)\citenamefont{{{\"O}zel},
  {Psaltis}, {Narayan}, and {McClintock}}}]{Ozel}
\bibinfo{author}{\bibfnamefont{F.}~\bibnamefont{{{\"O}zel}}},
  \bibinfo{author}{\bibfnamefont{D.}~\bibnamefont{{Psaltis}}},
  \bibinfo{author}{\bibfnamefont{R.}~\bibnamefont{{Narayan}}},
  \bibnamefont{and} \bibinfo{author}{\bibfnamefont{J.~E.}
  \bibnamefont{{McClintock}}}, \bibinfo{journal}{Astrophys. J.}
  \textbf{\bibinfo{volume}{725}}, \bibinfo{pages}{1918} (\bibinfo{year}{2010}),
  \eprint{1006.2834}.

\bibitem[{\citenamefont{{Bailyn} et~al.}(1998)\citenamefont{{Bailyn}, {Jain},
  {Coppi}, and {Orosz}}}]{Bailyn}
\bibinfo{author}{\bibfnamefont{C.~D.} \bibnamefont{{Bailyn}}},
  \bibinfo{author}{\bibfnamefont{R.~K.} \bibnamefont{{Jain}}},
  \bibinfo{author}{\bibfnamefont{P.}~\bibnamefont{{Coppi}}}, \bibnamefont{and}
  \bibinfo{author}{\bibfnamefont{J.~A.} \bibnamefont{{Orosz}}},
  \bibinfo{journal}{Astrophys. J.} \textbf{\bibinfo{volume}{499}},
  \bibinfo{pages}{367} (\bibinfo{year}{1998}), \eprint{astro-ph/9708032}.

\bibitem[{\citenamefont{{Fryer} et~al.}(2001)\citenamefont{{Fryer}, {Woosley},
  and {Heger}}}]{Fryer2001}
\bibinfo{author}{\bibfnamefont{C.~L.} \bibnamefont{{Fryer}}},
  \bibinfo{author}{\bibfnamefont{S.~E.} \bibnamefont{{Woosley}}},
  \bibnamefont{and} \bibinfo{author}{\bibfnamefont{A.}~\bibnamefont{{Heger}}},
  \bibinfo{journal}{apj} \textbf{\bibinfo{volume}{550}}, \bibinfo{pages}{372}
  (\bibinfo{year}{2001}), \eprint{astro-ph/0007176}.

\bibitem[{\citenamefont{{Yusof} et~al.}(2013)\citenamefont{{Yusof}, {Hirschi},
  {Meynet}, {Crowther}, {Ekstr{\"o}m}, {Frischknecht}, {Georgy}, {Abu Kassim},
  and {Schnurr}}}]{Yusof2013}
\bibinfo{author}{\bibfnamefont{N.}~\bibnamefont{{Yusof}}},
  \bibinfo{author}{\bibfnamefont{R.}~\bibnamefont{{Hirschi}}},
  \bibinfo{author}{\bibfnamefont{G.}~\bibnamefont{{Meynet}}},
  \bibinfo{author}{\bibfnamefont{P.~A.} \bibnamefont{{Crowther}}},
  \bibinfo{author}{\bibfnamefont{S.}~\bibnamefont{{Ekstr{\"o}m}}},
  \bibinfo{author}{\bibfnamefont{U.}~\bibnamefont{{Frischknecht}}},
  \bibinfo{author}{\bibfnamefont{C.}~\bibnamefont{{Georgy}}},
  \bibinfo{author}{\bibfnamefont{H.}~\bibnamefont{{Abu Kassim}}},
  \bibnamefont{and}
  \bibinfo{author}{\bibfnamefont{O.}~\bibnamefont{{Schnurr}}},
  \bibinfo{journal}{Mon. Not. Roy. Astron. Soc.}
  \textbf{\bibinfo{volume}{433}}, \bibinfo{pages}{1114} (\bibinfo{year}{2013}),
  \eprint{1305.2099}.

\bibitem[{\citenamefont{{Belczynski} et~al.}(2014)\citenamefont{{Belczynski},
  {Buonanno}, {Cantiello}, {Fryer}, {Holz}, {Mandel}, {Miller}, and
  {Walczak}}}]{Belczynski2014}
\bibinfo{author}{\bibfnamefont{K.}~\bibnamefont{{Belczynski}}},
  \bibinfo{author}{\bibfnamefont{A.}~\bibnamefont{{Buonanno}}},
  \bibinfo{author}{\bibfnamefont{M.}~\bibnamefont{{Cantiello}}},
  \bibinfo{author}{\bibfnamefont{C.~L.} \bibnamefont{{Fryer}}},
  \bibinfo{author}{\bibfnamefont{D.~E.} \bibnamefont{{Holz}}},
  \bibinfo{author}{\bibfnamefont{I.}~\bibnamefont{{Mandel}}},
  \bibinfo{author}{\bibfnamefont{M.~C.} \bibnamefont{{Miller}}},
  \bibnamefont{and}
  \bibinfo{author}{\bibfnamefont{M.}~\bibnamefont{{Walczak}}},
  \bibinfo{journal}{Astrophys. J.} \textbf{\bibinfo{volume}{789}},
  \bibinfo{eid}{120} (\bibinfo{year}{2014}), \eprint{1403.0677}.

\bibitem[{\citenamefont{{Kroupa} and {Weidner}}(2003)}]{Kroupa2003}
\bibinfo{author}{\bibfnamefont{P.}~\bibnamefont{{Kroupa}}} \bibnamefont{and}
  \bibinfo{author}{\bibfnamefont{C.}~\bibnamefont{{Weidner}}},
  \bibinfo{journal}{\apj} \textbf{\bibinfo{volume}{598}}, \bibinfo{pages}{1076}
  (\bibinfo{year}{2003}), \eprint{astro-ph/0308356}.

\end{thebibliography}

\end{document}